\documentclass[12pt,article,nofootinbib]{revtex4}%
\usepackage{amsfonts}
\usepackage{amsmath}
\usepackage{amssymb}
\usepackage{graphicx}
\usepackage{subfig}%
\providecommand{\U}[1]{\protect\rule{.1in}{.1in}}
\renewcommand{\thesection}{\arabic{section}}

\makeatletter \@addtoreset{equation}{section}
\renewcommand{\theequation}{\thesection.\arabic{equation}}
\begin{document}
\preprint{ }
\title{\rightline{\mbox{\small {LPHE-MS-18-03}} \vspace
{1cm}} \textbf{Hybrid seesaw neutrino model in SUSY }$SU(5)\times
\mathbb{A}_{4}$\textbf{\ }}
\author{R. Ahl Laamara, M. A. Loualidi, M. Miskaoui, and E. H. Saidi\thanks{E-mail:
h-saidi@fsr.ac.ma}}
\affiliation{{\small LPHE-Modeling and Simulations, Faculty of \ Sciences,}}
\affiliation{{\small Mohammed V University, Rabat, Morocco}}
\affiliation{{\small Center of Physics and Mathematics, CPM- Morocco}}

\begin{abstract}
Motivated by recent results from neutrino experiments, we study the neutrino
masses and mixing in the framework of a SUSY $SU(5)\times\mathbb{A}_{4}%
$\ model. The hybrid of Type I\textrm{\ }and Type II seesaw mechanisms leads
to the nonzero value of the reactor angle $\theta_{13}\neq0$\ and to the
recently disfavored maximal atmospheric angle\textrm{\ }$\theta_{23}%
\neq45^{\circ}$ by the NOvA experiment\textrm{.} The phenomenological
consequences of the model are studied for both normal and inverted mass
hierarchies. The obtained ranges for the effective Majorana neutrino mass
$m_{\beta\beta}$, the electron neutrino mass $m_{\nu_{e}}$, and the $CP$
violating phase $\delta_{CP}$ lie within the current experimental allowed
ranges where\textrm{\ }we find that the normal mass hierarchy is favored over
the inverted one.\newline\emph{Key words}: Neutrinos mixing, SUSY
$SU(5)\times\mathbb{A}_{4}$, Hybrid Seesaw.

\end{abstract}
\email{E-mail: h-saidi@fsr.ac.ma}
\volumeyear{ }
\volumenumber{ }
\issuenumber{ }
\eid{ }
\maketitle

%\tableofcontents

\section{Introduction}

The neutrino oscillation experiments performed in the past two decades
provided many decisive evidences of nonzero neutrino masses and large neutrino
mixing \cite{R1,R2,R3,R4,R5,A1}. The atmospheric, solar, and reactor neutrino
experiments have provided the measurements of the mass-squared differences
$\Delta m_{ij}^{2}$ as well as the mixing angles $\theta_{ij}$;\textrm{\ }the
current neutrino oscillation data can be found in the latest global fit
analysis \cite{R6,R7,R8}. To understand the origin of these masses\textbf{---}%
which are very tiny---and mixing, we must go beyond the standard model (SM)
that predicts massless neutrinos. Theoretically, the most prominent way to
generate such tiny masses for neutrinos is through the famous seesaw
mechanism, which requires the introduction of extra heavy fermions (Type I and
Type III seesaws) or scalars (Type II seesaw)\ into the SM \cite{R9,R10},
giving rise to neutrino masses of Majorana type. For the neutrino mixing
angles, it was not until 2012 that the reactor angle $\theta_{13}$\ was
discovered to be different from zero \cite{R3}, but unlike the other two
mixing angles $\theta_{12}$ and $\theta_{23}$, its value is relatively small.
Furthermore, the NOvA experiment has disfavored recently the maximal
atmospheric neutrino mixing $\sin^{2}\theta_{23}=0.5$\ \cite{cc}; however,
whether its value is less or greater than $\pi/4$\ is yet to be discovered. In
the Pontecorvo-Maki-Nakagawa-Sakata (PMNS) matrix that describes these angles,
$\theta_{13}$ always appears in combination with the Dirac phase, and thus,
the discovery of its nonzero value has a crucial influence on the Dirac $CP$
violating (CPV) phase $\delta_{CP}$ where its measurement is the ultimate
objective of the long baseline neutrino oscillation experiments \cite{R11}.
The recent progress in neutrino physics motivated theoretical as well as
experimental physicists to search for new physics beyond the SM. This concerns
the preexisting theories and models such as supersymmetric grand unified
theories (SUSY GUTs) which unlike the non-SUSY GUTs solve the hierarchy
problem and unification of gauge couplings just by introducing supersymmetry;
thus, they are adopted as one of the most appealing extensions of the SM
\cite{NAN}. Moreover, an attractive way to outline the observed neutrino mass
hierarchies and mixing within SUSY-GUT models is through discrete flavor
symmetries. Indeed, several models beyond SM have used different non-Abelian
groups and described successfully all the neutrino mixing angles; see Table 3
of Ref. \cite{R12} and Ref. \cite{R13}. In fact, these non-Abelian discrete
groups are widely adopted to describe the large mixing angles in the lepton
sector. In particular, these groups lead to a specific form of the neutrino
mass matrix which is consistent with tribimaximal mixing (TBM). This special
mixing induces $\theta_{13}=0$\ and $\theta_{23}=\pi/4$; however, it is now
ruled out by the discovery of the nonzero reactor angle as mentioned above.
Thus a small deviation from TBM is required to reconcile with the small value
of $\theta_{13}$ as well as a small deviation from the maximal value of the
atmospheric angle $\theta_{23}$. In this regard, several ways have been
proposed to generate a small deviation of these mixing angles. For example,
the deviation from TBM in flavor symmetry-based models can arise from (i) the
diagonalization of the charged lepton mass matrix \cite{DT}, (ii) perturbing
the vacuum expectation value (VEV) alignment \cite{VA}, (iii) the Yukawa
sector \cite{YS}, or (iv) the Majorana sector \cite{R13,MS}. These deviations
are generally realized by introducing\ next-to-leading-order effective
operators while the leading contribution is produced by one of the seesaw
mechanisms. On the other hand, it was claimed in Ref. \cite{HS} that the
required deviations from the TBM matrix can be interpreted as the interplay of
two different seesaw mechanisms making what is known as hybrid neutrino
masses. This hybrid has been used by many authors to account for the nonzero
reactor angle $\theta_{13}\neq0$ in the framework of the SM and GUTs; see, for
example, Ref. \cite{NM}.

In this paper we propose a neutrino model in the framework of a supersymmetric
$SU(5)$ GUT extended by three right-handed neutrinos $N_{i}$ and a
15-dimensional Higgs $H_{15}$ transforming respectively as a triplet and a
nontrivial singlet under $\mathbb{A}_{4}$ flavor symmetry. The theoretical
predictions of our proposal concerning the mixing angles and masses are
compatible with the latest neutrino experimental data. The main line of our
proposal is as follows: First, we consider SUSY $SU(5)\times$ $\mathbb{A}_{4}%
$\ theory and generate the neutrino mass matrix by the hybrid seesaw
mechanism. In this hybrid, the dominant mass contribution comes from Type I
seesaw, leading to the TBM \cite{HPS}. A small perturbation responsible for
nonzero reactor angle $\theta_{13}$ and nonmaximal atmospheric angle
$\theta_{23}$ is realized by the 15-dimensional $SU(5)$ Higgs that contains an
$SU(2)_{L}$ Higgs triplet $\Delta_{d}$ via Type II seesaw mechanism. Then, we
perform a numerical study, where we use the experimental allowed ranges of the
mixing angles and the mass-squared differences, to examine the octant
degeneracy of $\theta_{23}$ for both normal and inverted mass
hierarchies\emph{.} Next, we use the current neutrino oscillation data as well
as the cosmological limit on the sum of neutrino masses to study the
phenomenological consequences of our proposal for both normal and inverted
mass hierarchies. We find that the allowed ranges of the effective Majorana
neutrino mass $m_{ee}$, the sum of neutrino masses $\sum\nolimits_{i=1}%
^{3}\left\vert m_{i}\right\vert $, the effective\textrm{\ }electron neutrino
mass $m_{\beta}$, and the Dirac CPV phase $\delta_{CP}$ are within the current
experimental data.\newline To perform this study, we use known results on SUSY
$SU\left(  5\right)  $ as well as properties of the alternating group
$\mathbb{A}_{4}$. This flavor symmetry is generally admitted as the most
natural and economical discrete group that captures the family symmetry as
motivated in the literature \cite{R14}. The discrete $\mathbb{A}_{4}$
possesses two generators $S$, $T$ and four irreducible representations that
can be labeled by their characters as $\mathbf{1}_{\left(  1,1\right)  },$
$\mathbf{1}_{\left(  1,\omega\right)  },$ $\mathbf{1}_{\left(  1,\omega
^{2}\right)  }$, and $\mathbf{3}_{\left(  -1,0\right)  }$. These four
representations, which are related to the $\mathbb{A}_{4}$ group order by the
formula $\mathbf{1}_{\left(  1,1\right)  }^{2}+\mathbf{1}_{\left(
1,\omega\right)  }^{2}+\mathbf{1}_{\left(  1,\omega^{2}\right)  }%
^{2}+\mathbf{3}_{\left(  -1,0\right)  }^{2}=12$, are also used to host the
matter and Higgs content of the SUSY $SU(5)\times$ $\mathbb{A}_{4}$\ proposal.
For general properties on $\mathbb{A}_{4}$ group representations and their
characters, see \cite{R15,R16}.\newline This paper is organized as follows. In
Sec. 2, we present the superfield content for the neutrino sector in SUSY
$SU(5)\times$ $\mathbb{A}_{4}$. Then, we study the Dirac and Majorana neutrino
mass matrices as well as the deviations of $\theta_{13}$\ and $\theta_{23}$
from their TBM values. In Sec. 3, we study the phenomenological implications
of the proposal and provide the predictions regarding the effective Majorana
mass $m_{ee}$,\textrm{\ }the effective mass $m_{\beta}$, the sum
$\sum\nolimits_{i=1}^{3}\left\vert m_{i}\right\vert $, and the CPV phase
$\delta_{CP}$. In Sec. 4, we give our conclusion. In order to make the paper
more self-contained we add Appendix A on the charged sector where we show that
a $U(1)$\ flavor symmetry is needed to control the couplings of the model. We
also add in the same appendix a brief discussion on the well-known dangerous
four- and five-dimensional operators leading to the rapid proton decay and
show how they are suppressed in our model due to the\ flavor symmetry.

\section{$SU(5)$ GUT with $\mathbb{A}_{4}$ flavor symmetry}

In this section, we first describe the superfield content of our
$SU(5)\times\mathbb{A}_{4}$ GUT proposal. Then, we use a hybrid seesaw
mechanism to study the deviation of the $\theta_{13}$\ and $\theta_{23}%
$\ angles in this proposal. After that, we study the mass-squared differences
as functions of the space parameters of the model and the $\theta_{23}$\ and
$\theta_{13}$\ mixing angles.

\subsection{Implementing $\mathbb{A}_{4}$ in neutrino sector}

In supersymmetric $SU(5)$ GUT, matter superfields are unified into two
irreducible representations of $SU(5)$ namely $\mathbf{10}_{m}^{i}$ and
$\mathbf{\bar{5}}_{m}^{i}$ where $i=1,2,3$ refers to the three possible
generations of matter. On the other hand, the Higgs doublets $H_{u}$ and
$H_{d}$ of the minimal supersymmetric standard model (MSSM) sit in
representations $\mathbf{5}_{H_{u}}=H_{5}$ and $\overline{\mathbf{5}}_{H_{d}%
}=H_{\overline{5}}$.\newline Here we focus our attention on the neutrino
sector\textrm{\ }in SUSY\textrm{\ }$SU(5)$ GUT promoted by an\textrm{\ }%
$\mathbb{A}_{4}$ flavor symmetry. Thus, we give only the superfield content
needed to generate the mass terms for the neutrinos. In our construction of
SUSY $SU(5)\times\mathbb{A}_{4}$ GUT, we proceed as follows:

(i) First, we extend the fermion sector of $SU(5)$ GUT by adding three
right-handed neutrinos $N_{i}$\ which are $SU(5)$ gauge singlets and sit
together in the $\mathbb{A}_{4}$ triplet $\mathbf{3}_{-1,0}$. These $N_{i}$'s
allow us to use the Type I seesaw formula $m_{\nu}^{I}=-m_{D}M_{R}^{-1}%
m_{D}^{T}$ to generate light neutrino masses. One $\mathbb{A}_{4}$ flavon
triplet superfield $\Phi$ is added to get a neutrino mass matrix $m_{\nu}^{I}$
consistent with the leading order TBM pattern. The addition of one flavon in
the neutrino sector is actually the minimal setup if we consider only the
four-dimensional $SU(5)\times\mathbb{A}_{4}$\ models that describe
successfully all the mixing angles. Some of these models that used at least
three flavon superfields in the neutrino sector are given in Ref. \cite{C0}.

(ii) Second, we extend the Higgs sector of SUSY $SU(5)$ GUT by adding a
$\mathbf{15}$-dimensional Higgs $\mathbf{15}_{\Delta_{d}}\equiv H_{15}$ which
contains a $Y=2$ $SU(2)_{L}$ Higgs triplet $\Delta_{d}$. This leads to a
Majorana mass matrix $M_{\nu}^{II}$\ via the Type II seesaw mechanism as
exhibited by the Yukawa coupling\textrm{\ }$\mathbf{\bar{5}}_{m}%
\otimes\mathbf{15}_{\Delta_{d}}\otimes\mathbf{\bar{5}}_{m}$. When added to
$m_{\nu}^{I}$,\ the matrix $M_{\nu}^{II}$ will play the role of a\emph{\ }%
perturbation inducing a deviation from the TBM values. Notice that $H_{15}$
has been first used in non-SUSY $SU\left(  5\right)  $ without flavor symmetry
to achieve the gauge coupling unification and the generation of tiny neutrino
masses \cite{A2}. Notice also that the deviation from TBM by Type II seesaw
mechanism with discrete flavor $\mathbb{A}_{4}$ has also been considered in SM
to reconcile with the experimental value of $\theta_{13}$ \cite{C1}.
In\textbf{\ }our SUSY $SU(5)\times\mathbb{A}_{4}$ proposal which extends this
approach to supersymmetric GUT models building, we took into account the
latest experimental results on neutrino masses and mixing, and we successfully
produced the nonzero value of $\theta_{13}$ as well as the nonmaximal value of
$\theta_{23}$.

\ \ \ \ \ \newline So the superfield content of our proposal is as follows:
\textrm{(a)} matter containing three generations of $\overline{\mathbf{5}}%
_{m}^{i}$ denoted as $F_{i}$, $\mathbf{10}_{m}^{i}$ denoted as $T_{i}$, and
the three right-handed neutrinos $N_{i}$. Below, we will mainly focus on
$F_{i}$ and $N_{i}$ couplings relevant for neutrino sector, while the
contribution of the $\mathbf{10}_{m}^{i}$'s in the charged lepton and quark
sectors will be discussed in Appendix A. \textrm{(b)} The Higgs sector
containing: \textrm{(i)} the two usual Higgses $H_{5}$\ and $H_{\bar{5}}$\ as
well as the added $H_{15}$ and $H_{\overline{15}}$; the $H_{\bar{5}}$\ and
$H_{\overline{15}}$\ are required by supersymmetry. \textrm{(ii)} The usual
24-dimensional adjoint Higgs $H_{24}$\ needed to break the $SU(5)$\ group to
the standard model gauge group. \textrm{(iii)} An extra flavon chiral
superfield $\Phi$ to generate the TBM matrix.\newline These superfields are
the minimal set we need to generate neutrino masses and mixing compatible with
experimental data. The quantum numbers of these superfields under
$SU(5)\times\mathbb{A}_{4}$ are as listed in Table \ref{t1}.

\begin{table}[h]
\centering \renewcommand{\arraystretch}{1.2} $%
\begin{tabular}
[c]{|l|l|l|l|l|l|l|l|l|l|}\hline
Fields & $F_{i}$ & $T_{1}$ & $T_{2}$ & $T_{3}$ & $N_{i}$ & $H_{5}$ &
$H_{\bar{5}}$ & $\Phi$ & $H_{15}$\\\hline
$\mathrm{SU(5)}$ & $\bar{5}_{m}^{i}$ & $10_{m}^{1}$ & $10_{m}^{2}$ &
$10_{m}^{3}$ & $1_{\nu}^{i}$ & $5_{H_{u}}$ & $5_{H_{d}}$ & $1$ &
$15_{\Delta_{d}}$\\\hline
$A_{4}$ & $3_{-1,0}$ & $1_{\left(  1,\omega\right)  }$ & $1_{\left(
1,\omega^{2}\right)  }$ & $1_{\left(  1,1\right)  }$ & $3_{-1,0}$ &
$1_{\left(  1,1\right)  }$ & $1_{\left(  1,\omega\right)  }$ & $3_{-1,0}$ &
$1_{\left(  1,\omega\right)  }$\\\hline
\end{tabular}
\ \ $ \caption{Superfield content and their quantum numbers under
$SU(5)\times\mathbb{A}_{4}$.}%
\label{t1}%
\end{table}Besides $N_{i}$ and $\Phi$,\ which are gauge singlets, $T_{i},$
$F_{i},$ $H_{5}$, and $H_{15}$ are given in standard model representations as
follows:%
\begin{equation}%
\begin{tabular}
[c]{lllll}%
$\mathrm{SU(5)}$ & $\ {\small \rightarrow}$ &
\multicolumn{3}{l}{$\ \ \ \ \ \ \ \ \ \mathrm{SU(3)}_{\mathrm{c}%
}\mathrm{\times SU(2)}_{\mathrm{L}}\mathrm{\times U(1)}_{\mathrm{Y}}$}\\\hline
$\ \mathbf{\bar{5}}_{m}^{i}$ & $\ :$ & $\ \ \left(  \bar{3},1\right)
_{2/3}+\left(  1,2\right)  _{-1}$ & ${\small =}$ & $\left(  D_{i}^{c}%
,L_{i}\right)  $\\
$\ \mathbf{5}_{H_{u}}$ & $\ :$ & $\ \ \left(  3,1\right)  _{-2/3}+\left(
1,2\right)  _{1}$ & ${\small =}$ & $\left(  T_{u},H_{u}\right)  $\\
$\ \mathbf{15}_{\Delta_{d}}$ & $\ :$ & $\ \ (1,3)_{2}+(3,2)_{1/3}%
+(6,1)_{-4/3}$ & ${\small =}$ & $\left(  \Delta_{d},\Delta_{d}^{\prime}%
,\Delta_{d}^{\prime\prime}\right)  $\\
$\mathbf{10}_{m}^{i}$ & $~:$ & $\ (3,2)_{1/3}+(\bar{3},1)_{-4/3}+(1,1)_{2}$ &
$=$ & $(Q_{i},U_{i}^{c},E_{i}^{c})$\\\hline
\end{tabular}
\ \ \ \ \ \ \ \ \label{d15}%
\end{equation}
where the decompositions of $5_{H_{d}}$\ and $\overline{15}_{\Delta_{u}}$\ are understood.

\subsection{Deviation of $\theta_{13}$\ and $\theta_{23}$\ in $SU(5)\times
\mathbb{A}_{4}$ hybrid seesaw}

We start with the leading approximation where the neutrino mass matrix is
generated through Type I seesaw mechanism and is consistent with TBM
predicting the mixing angles: $\sin^{2}\theta_{12}=\frac{1}{3}$, $\sin
^{2}\theta_{23}=\frac{1}{2}$, and $\sin^{2}\theta_{13}=0$. Then, we make use
of the 15-dimensional $SU(5)$ Higgs $\mathbf{15}_{\Delta_{d}}$ that
contains\textrm{\ }an $SU\left(  2\right)  _{L}$ Higgs triplet $\Delta_{d}$
leading to Majorana mass term via Type II seesaw mechanism. Hence, the total
neutrino mass matrix combines both Type I and Type II seesaws, allowing a
reconciliation with the experimental values of the mixing angles $\theta_{13}%
$\ and $\theta_{23}$.

\subsubsection{TBM from Type I seesaw mechanism}

The Type I seesaw formula incorporates both Dirac and Majorana mass matrices
where the Dirac mass matrix $m_{D}$\ is obtained from the superpotential term
involving the couplings among the superfields $N_{i}$, $F_{i}$, and $H_{5}$
while the Majorana mass matrix $M_{R}$\ is obtained from the superpotential
involving the coupling of right-handed neutrinos $N_{i}$ with themselves. As
we mentioned before, both $F_{i}$ and $N_{i}$ live in the $\mathbb{A}_{4}%
$\ triplet $3_{-1,0}$ while the Higgs $H_{5}$\ is assigned to the trivial
singlet. The leading order superpotential for neutrino Yukawa couplings
respecting gauge and $\mathbb{A}_{4}$ symmetries is given by
\begin{equation}
W_{D}=\lambda_{1}NFH_{5}, \label{De}%
\end{equation}
where $\lambda_{1}$ is a Yukawa coupling constant. Using the tensor product of
$\mathbb{A}_{4}$\ irreducible representations in the Altarelli-Feruglio basis
\cite{R15, B0}, the superpotential (\ref{De}) reads%
\begin{equation}
W_{D}=\lambda_{1}\left(  N_{1}F_{1}H_{5}+N_{2}F_{3}H_{5}+N_{3}F_{2}%
H_{5}\right)  .
\end{equation}
When the Higgs doublet develops its VEV as the usual $\left\langle
H_{u}\right\rangle =\upsilon_{u}$, we get the Dirac mass matrix of neutrinos
as%
\begin{equation}
m_{D}=\upsilon_{u}\left(
\begin{array}
[c]{ccc}%
\lambda_{1} & 0 & 0\\
0 & 0 & \lambda_{1}\\
0 & \lambda_{1} & 0
\end{array}
\right)  . \label{md}%
\end{equation}
As for the Majorana mass matrix, the superpotential respecting gauge and
flavor symmetries of our model are given by%
\begin{equation}
W_{R}=m_{R}NN+\lambda_{2}NN\Phi, \label{wm}%
\end{equation}
where we have added the second term involving the flavon$\ \Phi$\ to satisfy
the TBM texture and to generate appropriate masses for the neutrinos. This
term---which is at the renormalizable level---will contribute to all the
entries in the Majorana mass matrix. By using the multiplication rules of
$\mathbb{A}_{4}$, the superpotential $W_{R}$ develops into%
\begin{equation}%
\begin{array}
[c]{ccc}%
W_{R} & = & m_{R}\left(  N_{1}N_{1}+N_{2}N_{3}+N_{3}N_{2}\right)
+\frac{\lambda_{2}}{3}\left(  2N_{1}N_{1}-N_{2}N_{3}-N_{3}N_{2}\right)
\Phi_{1}\\
&  & +\frac{\lambda_{2}}{3}\left(  2N_{3}N_{3}-N_{1}N_{2}-N_{2}N_{1}\right)
\Phi_{3}+\frac{\lambda_{2}}{3}\left(  2N_{2}N_{2}-N_{1}N_{3}-N_{3}%
N_{1}\right)  \Phi_{2}%
\end{array}
\end{equation}
and by taking the VEV of the flavons $\Phi$\ as $\left\langle \Phi
_{1}\right\rangle =\left\langle \Phi_{2}\right\rangle =\left\langle \Phi
_{3}\right\rangle =\upsilon_{\Phi}$, we find the Majorana neutrino mass matrix
$M_{R}$ given by%
\begin{equation}
M_{R}=m_{R}\left(
\begin{array}
[c]{ccc}%
1+2\alpha & -\alpha & -\alpha\\
-\alpha & 2\alpha & 1-\alpha\\
-\alpha & 1-\alpha & 2\alpha
\end{array}
\right)  \quad\text{with\quad}\alpha=\frac{\lambda_{2}\upsilon_{\Phi}}{3m_{R}%
}. \label{maj}%
\end{equation}
The light neutrino mass matrix is obtained using\textbf{\ }Type I seesaw
mechanism formula $m_{\nu}^{I}=-m_{D}M_{R}^{-1}m_{D}^{T}$ with the Dirac mass
matrix as in Eq. (\ref{md}), and we find%
\begin{equation}
m_{\nu}^{I}=-m_{0}\left(
\begin{array}
[c]{ccc}%
a & b & b\\
b & c & a+b-c\\
b & a+b-c & c
\end{array}
\right)  , \label{mn}%
\end{equation}
$\allowbreak$where we have adopted the following parametrization%
\begin{equation}
a=\frac{\alpha+1}{3\alpha+1}\quad,\quad b=\frac{\alpha}{3\alpha+1}\quad,\quad
c=\frac{3\alpha^{2}+2\alpha}{9\alpha^{2}-1}\quad,\quad m_{0}=\frac{\lambda
_{1}^{2}\upsilon_{u}^{2}}{m_{R}}. \label{pp}%
\end{equation}
Moreover, the values of the parameters $a$ and $b$ are related as $a=1-2b$;
this property will be used in our numerical study. The matrix $m_{\nu}^{I}$
respects the well-known $\mu-\tau$\ reflection symmetry \cite{R18}, and the
condition among the elements $\left(  m_{\nu}^{I}\right)  _{11}+\left(
m_{\nu}^{I}\right)  _{12}=\left(  m_{\nu}^{I}\right)  _{22}+\left(  m_{\nu
}^{I}\right)  _{23}$\ required to diagonalize $m_{\nu}^{I}$\ by the TBM matrix
as $m_{\nu}^{I}=U_{\mathrm{TBM}}^{T}m_{\nu}U_{\mathrm{TBM}}=diag(m_{1}%
,m_{2},m_{3})$\ where the $U_{\mathrm{TBM}}$\ is given by%
\begin{equation}
U_{TBM}=\left(
\begin{array}
[c]{ccc}%
-\sqrt{2/3} & 1/\sqrt{3} & 0\\
1\sqrt{6} & 1/\sqrt{3} & -1/\sqrt{2}\\
1/\sqrt{6} & 1/\sqrt{3} & 1/\sqrt{2}%
\end{array}
\right)  .
\end{equation}

\subsubsection{Deviation using Type II seesaw mechanism\textbf{\ }}

Now we turn to study the deviation from TBM, which consists of inducing a
small perturbation in the neutrino mass matrix. This deviation is motivated by
the fact that the current experimental data on solar and atmospheric mixing
angles are inadequate with the TBM values. The current $3\sigma$\ ranges of
the three mixing angles obtained from the global analysis in Ref. \cite{R8}
are given by%
\begin{align}
0.271  &  \leq\sin^{2}\theta_{12}\leq0.345,\nonumber\\
0.385(0.393)  &  \leq\sin^{2}\theta_{23}\leq0.635(0.640),\label{mi}\\
0.01934(0.01953)  &  \leq\sin^{2}\theta_{13}\leq0.02393(0.02408)\nonumber
\end{align}
for a normal (inverted) mass hierarchy.\textrm{ }As mentioned
above,\textrm{\ }the perturbation is carried out through Type II seesaw, which
implies the introduction of a scalar $SU\left(  2\right)  _{L}$ triplet
$\Delta_{d}$\ belonging to the 15-dimensional representation $H_{15}$ of the
$SU(5)$ gauge group. The $SU(5)\times\mathbb{A}_{4}$-invariant\ superpotential
induces the Yukawa coupling involving $\Delta_{d}$ as%
\begin{equation}
W^{II}=\lambda_{3}\overline{5}_{m}\overline{5}_{m}15_{\Delta_{d}}=\lambda
_{3}FFH_{15}.
\end{equation}
Using the VEV $\left\langle \Delta_{d}\right\rangle =\upsilon_{\Delta_{d}}$ of
the $SU(2)_{L}$\ triplet component of $H_{15}\equiv15_{\Delta_{d}}$, the
Majorana neutrino mass matrix reads as follows:
\begin{equation}
M_{\nu}^{II}=m_{0}\left(
\begin{array}
[c]{ccc}%
0 & 0 & \mathrm{\varepsilon}\\
0 & \mathrm{\varepsilon} & 0\\
\mathrm{\varepsilon} & 0 & 0
\end{array}
\right)  \quad\text{with\quad}\mathrm{\varepsilon}=\lambda_{3}\frac
{\upsilon_{\Delta_{d}}}{m_{0}}, \label{s2}%
\end{equation}
where we factored this matrix by $m_{0}$ to form a dimensionless deviation
parameter $\mathrm{\varepsilon}$\ as well as to ease the hybridization between
the seesaw mechanisms. Even though the tiny mass of neutrinos is encoded in
the VEV of the Higgs triplet---which is expressed as the ratio of the Higgs
doublets VEVs and the Higgs triplet mass \cite{R10}---in ordinary seesaw Type
II models, in the present paper we will discuss its contribution only through
the deviation parameter $\varepsilon$\ as we will see later when we perform a
numerical study concerning the oscillation parameters. In addition, it is well
known that the phenomenological constraint from the $\rho$ parameter that
measures the ratio between the neutral and charged currents \cite{B2}
restricts the VEVs of the Higgs multiplets higher than dimension two
\cite{B3}. As in our model the calculation of the $\rho$ parameter requires
taking into consideration at least three kinds of Higgs superfields---namely
an $SU(2)$\ triplet that belongs to $15_{\Delta_{d}}$\ with hypercharge $Y=2$,
an $SU(2)$\ triplet that belongs to $\overline{15}_{\Delta_{u}}$\ with $Y=-2$,
and an $SU(2)$\ triplet that belongs to $24_{H}$\ with $Y=0$---we leave
detailed investigations to future work.

Now, we turn to the total neutrino mass matrix generated by the hybrid seesaw
mechanism that consists of combining the contribution of Type II seesaw in Eq.
(\ref{s2}) and the one arisen from the Type I seesaw in Eq. (\ref{mn}) as
$m_{\nu}=m_{\nu}^{I}+M_{\nu}^{II}$ with%
\begin{equation}
m_{\nu}=m_{0}\left(
\begin{array}
[c]{ccc}%
-a & -b & \mathrm{\varepsilon}-b\\
-b & \mathrm{\varepsilon}-c & c-b-a\\
\mathrm{\varepsilon}-b & c-b-a & -c
\end{array}
\right)  , \label{gm}%
\end{equation}
where $a,$ $b$, and $c$ are as given in Eq. (\ref{pp}). The neutrino mass
matrix is diagonalized by a transformation such as $m_{\nu}^{\mathrm{diag}%
}=\tilde{U}^{T}m_{\nu}\tilde{U}$ where the system of eigenvectors and
eigenvalues can be developed as power series of $\mathrm{\varepsilon}$; we
find up to order $\mathcal{O}(\mathrm{\varepsilon}^{2})$, the matrix
$\tilde{U}$ given in terms of its eigenvectors as%
\begin{equation}
\tilde{U}=\left(
\begin{array}
[c]{ccc}%
-\sqrt{\frac{2}{3}} & \frac{1}{\sqrt{3}} & \frac{\mathrm{\varepsilon}}%
{2\sqrt{2}\left(  a-c\right)  }\\
\frac{1}{\sqrt{6}}-\frac{\sqrt{3}\mathrm{\varepsilon}}{4\sqrt{2}\left(
a-c\right)  } & \frac{1}{\sqrt{3}} & -\frac{1}{\sqrt{2}}-\frac
{\mathrm{\varepsilon}}{4\sqrt{2}\left(  a-c\right)  }\\
\frac{1}{\sqrt{6}}+\frac{\sqrt{3}\mathrm{\varepsilon}}{4\sqrt{2}\left(
a-c\right)  } & \frac{1}{\sqrt{3}} & \frac{1}{\sqrt{2}}-\frac
{\mathrm{\varepsilon}}{4\sqrt{2}\left(  a-c\right)  }%
\end{array}
\right)  \text{ }+\mathcal{O}\left(  \mathrm{\varepsilon}^{2}\right)
\label{dm}%
\end{equation}
and eigenvalues
\begin{equation}
m_{1}=m_{0}\left(  b-a-\frac{\mathrm{\varepsilon}}{2}\right)  \quad,\quad
m_{2}=-m_{0}\left(  a+2b-\mathrm{\varepsilon}\right)  \quad,\quad m_{3}%
=m_{0}\left(  b+a-2c+\frac{\mathrm{\varepsilon}}{2}\right)  \label{mm}%
\end{equation}
Consequently, the mixing angles $\theta_{13}$\ and $\theta_{23}$\ become
\begin{equation}
\sin\theta_{13}=\left\vert \frac{\mathrm{\varepsilon}}{2\sqrt{2}\left(
a-c\right)  }\right\vert \text{\quad},\text{\quad}\sin\theta_{23}=\left\vert
\frac{\mathrm{\varepsilon}}{4\sqrt{2}\left(  a-c\right)  }+\frac{1}{\sqrt{2}%
}\right\vert \label{ss}%
\end{equation}
while the solar angle maintains its TBM (maximal) value $\sin\theta
_{12}=1/\sqrt{3}$. We have now a nonvanishing reactor angle $\theta_{13}$\ and
a small shift from the TBM value for the atmospheric angle $\theta_{23}$.

\subsection{Mass-squared differences and mixing angles}

\begin{figure}[ptbh]
\begin{center}
\hspace{1.7cm} \includegraphics[scale=1.1]{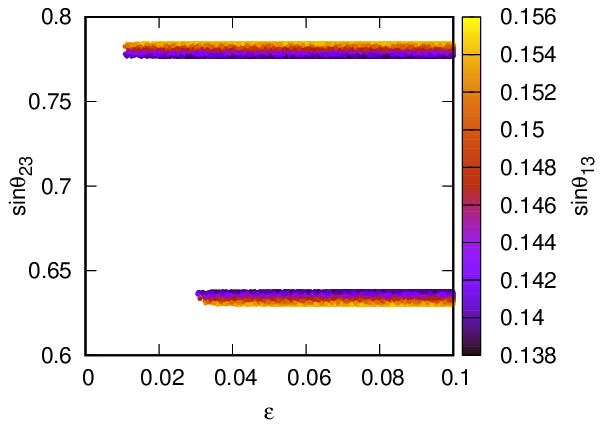}
\includegraphics[scale=1.1]{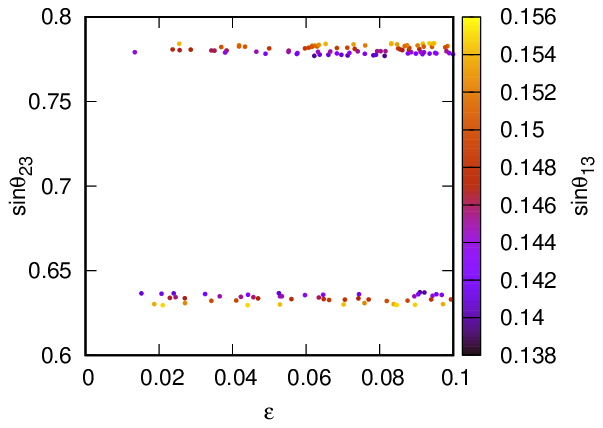}
\end{center}
\par
\vspace{1.5cm} \caption{$\sin\theta_{23}$ as a function of the parameter of
deviation $\varepsilon$ with $\sin\theta_{13}$\ shown in the palette for the
NH (left panel) and the IH (right panel).}%
\label{01}%
\end{figure}

\begin{figure}[ptbh]
\begin{center}
\hspace{2.5em}\includegraphics[scale=0.54]{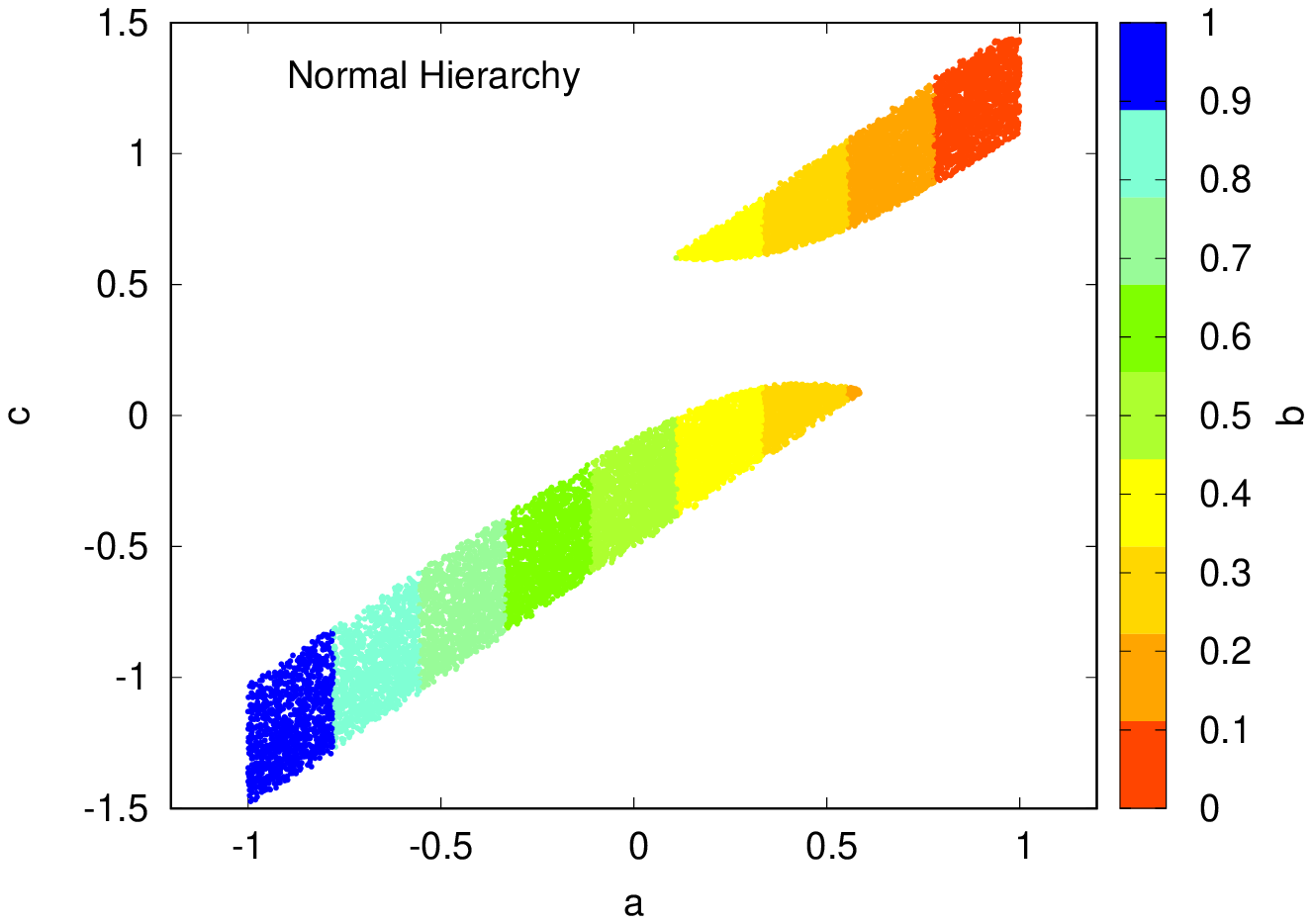}\quad
\includegraphics[scale=0.54]{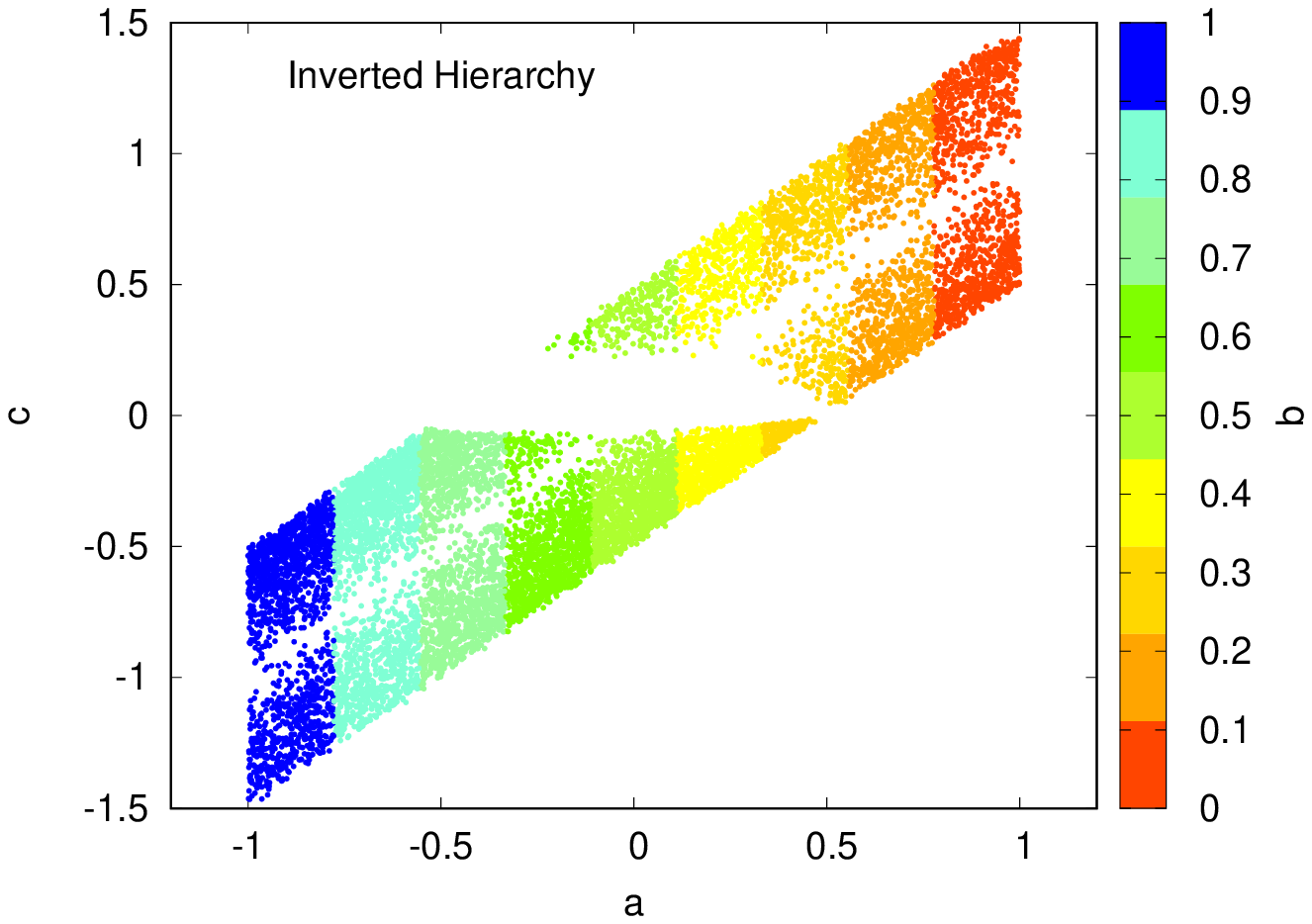}
\end{center}
\par
\vspace{0.8cm} \caption{Correlation among the parameters $a$, $b$, and $c$.}%
\label{03}%
\end{figure}

Concerning neutrino masses, the current neutrino oscillation experiments are
only sensitive to mass-squared differences where we distinguish between two
mass hierarchies: normal mass hierarchy (NH) where $m_{1}<m_{2}<m_{3}$ and
inverted mass hierarchy (IH) where $m_{3}<m_{1}<m_{2}$. Their $3\sigma$
experimental ranges are given by \cite{R8}%
\begin{align}
0.0000703  &  \leq\Delta m_{21}^{2}\leq0.0000809,\nonumber\\
(0.002399)0.002407  &  \leq\left\vert \Delta m_{3l}^{2}\right\vert
\leq0.002643(0.002635) \label{dv}%
\end{align}
with $l=1$ ($l=2$) for NH (IH). In our proposal, by using the masses in Eq.
(\ref{mm}), the solar $\Delta m_{21}^{2}$ and atmospheric $\Delta m_{3l}^{2}%
$\ mass-squared differences up to first order in $\varepsilon$ are expressed
as
\begin{align}
\Delta m_{21}^{2}  &  =3m_{0}^{2}\left(  b^{2}-b\mathrm{\varepsilon
}+2ab-a\mathrm{\varepsilon}\right)  ,\nonumber\\
\Delta m_{31}^{2}  &  =2m_{0}^{2}\left(  b-c\right)  \left(
2a-2c+\mathrm{\varepsilon}\right)  ,\label{dg}\\
\Delta m_{32}^{2}  &  =m_{0}^{2}\left(  3a\mathrm{\varepsilon}%
-2ab-2c\mathrm{\varepsilon}-3b^{2}+5b\mathrm{\varepsilon}-4c\left(
a+b-c\right)  \right)  .\nonumber
\end{align}
By using the mixing angles in Eq. (\ref{ss}), we show in the left panel (right
panel) of Fig. \ref{01} the correlation among the parameters $\sin\theta_{23}%
$, $\varepsilon$, and $\sin\theta_{13}$ for the NH case (IH case). The
experimental inputs of the mass squared differences $\Delta m_{31}^{2}%
$($\Delta m_{32}^{2}$) as well as their expressions given in Eq. (\ref{dg})
are taken into account. Before we discuss the ranges of the oscillation parameters, we
should notice that the recent measurement of the atmospheric angle\ from the
NOvA experiment disfavored the maximal value $\theta_{23}=45^{\circ}$
\cite{cc}, while experiments like T2K \cite{R5}\textrm{\ }and IceCube
\cite{B1} still prefer maximal mixing. In the case of nonmaximal mixing, there
are two different octants of $\theta_{23}$; the lower octant (LO) with
$\theta_{23}<45^{\circ}$ and the higher octant (HO) with $\theta
_{23}>45^{\circ}$. The NOvA experiment provided two degenerate ranges for the
normal mass hierarchy \cite{cc}: $\sin^{2}\theta_{23}=0.404_{-0.022}^{+0.030}$
(LO) and $\sin^{2}\theta_{23}=0.624_{-0.030}^{+0.022}$ (HO).\emph{\ }Back to
Fig. \ref{01}, we observe that while the entire $3\sigma$ range of $\sin
\theta_{13}$ is allowed, the ranges of the atmospheric angle become more
restrained. In the left panel (normal hierarchy), we observe that both octants
of the atmospheric angle are allowed and we have\textrm{ }
\begin{equation}
0.629\lesssim\sin\theta_{23}\text{(LO)}\lesssim0.637\quad,\quad0.776\lesssim
\sin\theta_{23}\text{(HO)}\lesssim0.784.
\end{equation}
These intervals correspond to
\begin{equation}
0.03\lesssim\mathrm{\varepsilon}\text{(LO)}\leq0.1\quad,\quad0.01\lesssim
\mathrm{\varepsilon}\text{(HO)}\leq0.1.
\end{equation}
In the right panel (inverted hierarchy), we have for both octants of the
atmospheric angle\textrm{ }
\begin{equation}
0.629\lesssim\sin\theta_{23}\text{(LO)}\lesssim0.637\quad,\quad0.777\lesssim
\sin\theta_{23}\text{(HO)}\lesssim0.784,
\end{equation}
which correspond to the following intervals of the deviation parameter:
\begin{equation}
0.015\lesssim\mathrm{\varepsilon}\text{(LO)}\leq0.1\quad,\quad0.013\lesssim
\mathrm{\varepsilon}\text{(HO)}\leq0.1.
\end{equation}
In our proposal, it is clear that the maximal atmospheric angle, which
corresponds to $\sin\theta_{23}\simeq0.7$ in both panels of\ Fig. \ref{01}, is
excluded. In fact, this is due to the contribution of the Higgs triplet
$\Delta_{d}\in\mathbf{15}_{\Delta_{d}}$\ (encoded in the parameter
$\varepsilon$) which led to the Majorana mass matrix\ (\ref{s2}) via Type II
seesaw mechanism, allowing us to explain the nonzero reactor angle
$\theta_{13}\neq0$\ as well as providing a deviation of the atmospheric angle
from its maximal value. All the allowed regions predicted in our model
for\ $\sin\theta_{23}$ in the case of normal hierarchy are within the ranges
of LO and HO provided by the NOvA experiment. To plot the above figures, we
have taken $\left\vert a\right\vert \lesssim1$ and $\left\vert b\right\vert
\lesssim1$ which is clear from Eq. (\ref{pp}) while the parameter $c$ is
allowed to vary freely. Moreover, as the parameter of deviation
$\mathrm{\varepsilon}$ has to be small, we have taken its range to be around
$\mathcal{O}(\frac{1}{10})$. We have also fixed $m_{0}$ in the range $\left[
0,\frac{1}{10}\right]  $ since it is well known that the mass of the
right-handed neutrinos---proportional to $m_{R}$---lies at a scale beyond the
reach of present experiments, and it is usually taken at the GUT scale in
grand unified theories. \newline As a follow-up to the above discussion, it is
clear that the intervals of the parameters $a$, $b$,\ and $c$---expressed as a
function of $\alpha=(\lambda_{2}\upsilon_{\Phi}/3m_{R})$---are fixed according
to Eq. (\ref{pp}).\ However, in order to find their restricted ranges
compatible with the oscillation experiments, we plot in Fig. \ref{03} the
correlation among them\ by using the $3\sigma$\ experimental values of the
mixing angles and the mass-squared differences as well as Eqs. (\ref{ss}) and
(\ref{dg}). Hence, we observe that for both mass hierarchies, the allowed
ranges for the parameter $b$ is around $[0,0.9997]$, while the parameters $a$
and $c$ vary in the ranges $[-1,1]$ and $[-1.47,1.43]$, respectively. These
new ranges will be used as inputs to perform a numerical study concerning the
phenomenology of neutrino in the next section. 

\section{Phenomenological implications}

In this section, by using the model parameters that are restricted by the
$3\sigma$ experimental values of the mixing angles and the mass-squared
differences, we show by means of scatter plots for both hierarchies the
physical observables $m_{ee}$ and $m_{\beta}$ related respectively to
neutrinoless double beta decay and tritium beta decay experiments, and we also
provide scatter plot predictions on the sum of neutrino masses as well as on
the Dirac $CP$ violating phase.

\subsection{Neutrinoless double beta decay}

One of the most known neutrino mass related experiments is the neutrinoless
double beta decay $(0\nu\beta\beta)$ process, which has not been observed yet.
Its discovery would prove that neutrinos are Majorana particles, and it would
also prove that the lepton number $L$ is violated. The decay amplitude\ for
the $0\nu\beta\beta$\ process is proportional to the effective Majorana
neutrino mass given by \cite{B4}%
\begin{equation}
\left\vert m_{ee}\right\vert =\left\vert \sum_{i=1}^{3}U_{ei}^{2}%
.m_{i}\right\vert , \label{me1}%
\end{equation}
where $m_{i}$ are the three neutrino masses and $U_{ei}$\ are the elements of
the first row of the PMNS matrix \cite{R23}\textrm{.} In our proposal, this
mixing matrix is given by%
\begin{equation}
\tilde{U}_{d}=\tilde{U}.\mathrm{diag}(1,e^{i\alpha},e^{i\beta}),
\end{equation}
where $\alpha$ and $\beta$\ are the Majorana $CP$ violating phases and
$\tilde{U}$ is given in Eq. (\ref{dm}). Currently, the most recent bounds of
$m_{ee}$ come from the KamLAND-Zen \cite{R24} and GERDA \cite{R25}
experiments; they are respectively given by%
\begin{equation}
\left\vert m_{ee}\right\vert <0.061-0.165\text{ }\mathrm{eV}\quad
,\quad\left\vert m_{ee}\right\vert <0.15-0.33\text{ }\mathrm{eV.} \label{mee}%
\end{equation}
To study the variation the effective Majorana mass $m_{ee}$\ with the lightest
neutrino mass in our model for both hierarchies, we replace $U_{ei}$\ in Eq.
(\ref{me1}) by the elements of the first row of $\tilde{U}_{d}$; the effective
Majorana mass takes the form%
\begin{equation}
\left\vert m_{ee}\right\vert =\left\vert \frac{2m_{1}}{3}+\frac{m_{2}}%
{3}e^{2i\alpha}+\frac{m_{3}}{8}\frac{\mathrm{\varepsilon}^{2}}{\left(
a-c\right)  ^{2}}e^{2i\beta}\right\vert . \label{ee}%
\end{equation}
Furthermore, for the NH case where $m_{1}$\ is the lightest neutrino mass, we
substitute $m_{2}$\ by $\sqrt{\Delta m_{21}^{2}+m_{1}^{2}}$ and $m_{3}$ by
$\sqrt{\Delta m_{31}^{2}+m_{1}^{2}}$, and for the IH case where $m_{3}$\ is
the lightest neutrino mass, we substitute $m_{2}$\ by $\sqrt{m_{3}^{2}-\Delta
m_{32}^{2}}$\ and $m_{1}$\ by\textrm{\ }$\sqrt{m_{3}^{2}-\Delta m_{32}%
^{2}-\Delta m_{21}^{2}}$. The explicit forms of $m_{i}$ and $\Delta m_{ij}%
^{2}$\ as a function of parameter space of the model are as shown in Eqs.
(\ref{mm}) and (\ref{dg}). \begin{figure}[th]
\begin{center}
\hspace{2.5em} \includegraphics[width=.44\textwidth]{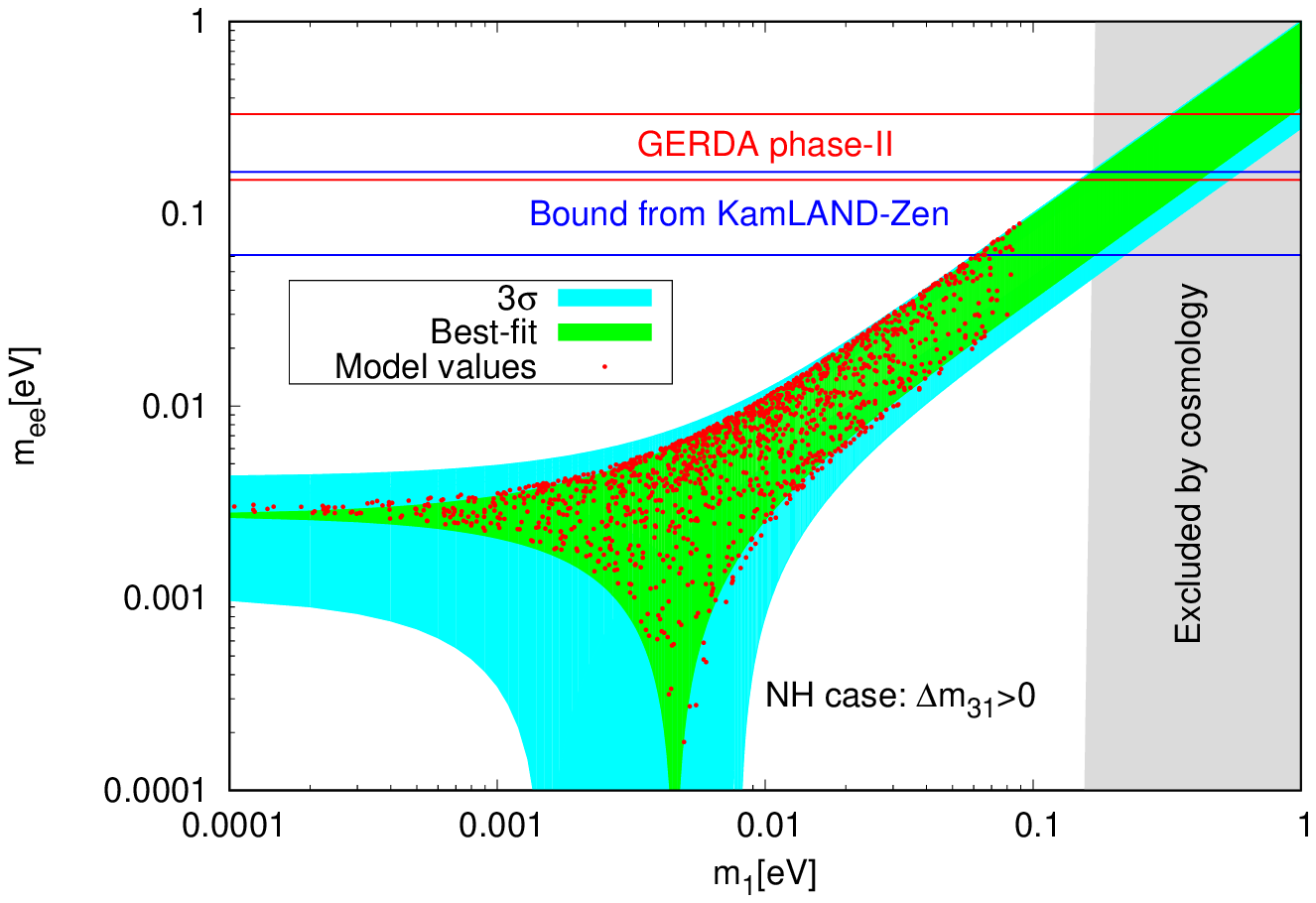}\quad
\includegraphics[width=.44\textwidth]{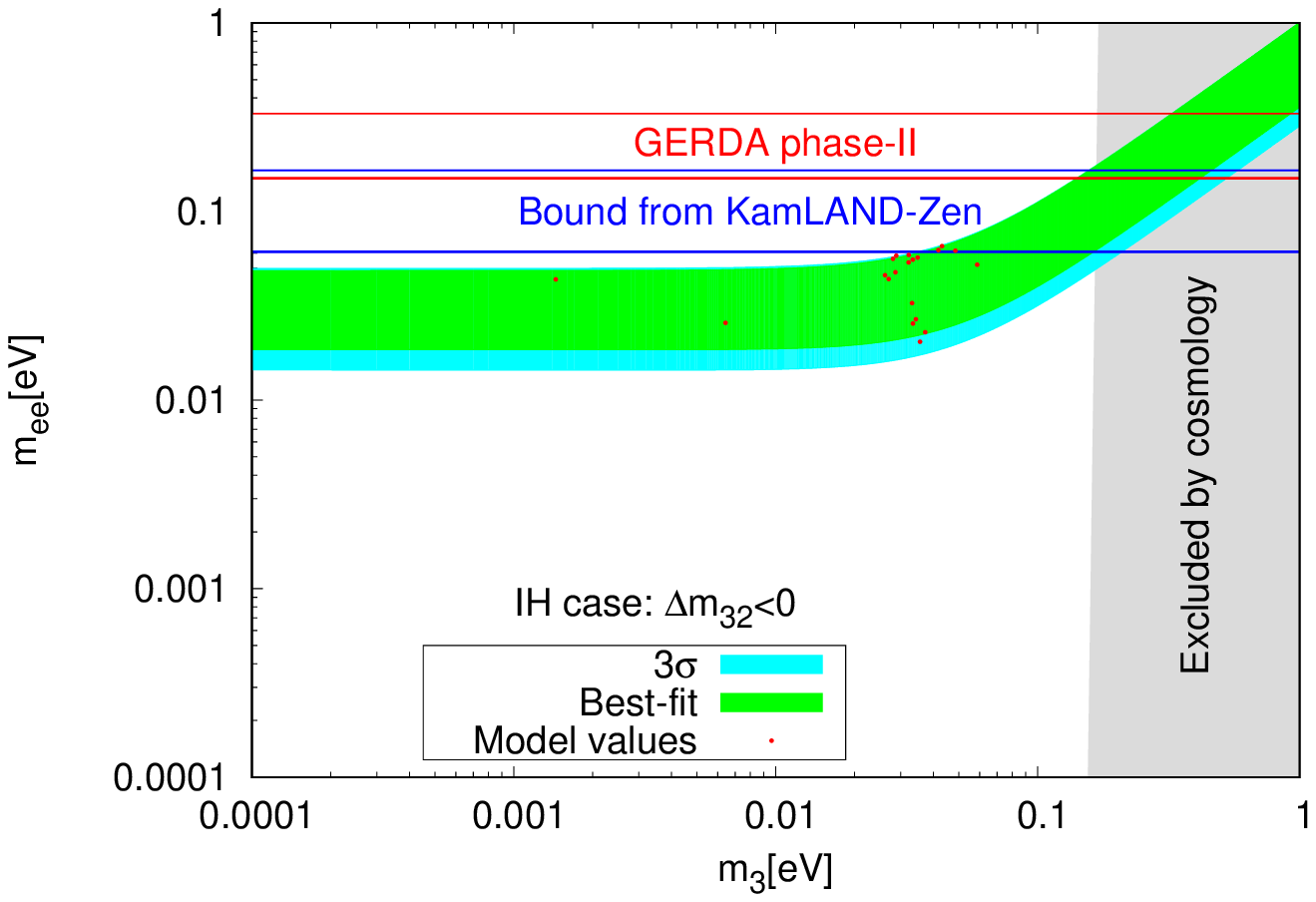}
\end{center}
\par
\vspace{0.5cm}\caption{Left: The effective Majorana mass as a function of the
lightest neutrino mass for NH. Right: Same as in the left panel but for IH.
The horizontal gray band in both panels indicates an upper limit on the sum of
the three light neutrino masses\ from Planck Collaboration.}%
\label{04}%
\end{figure}By using the above definitions and the limits from
experiments---see Eq. (\ref{mee})---we plot in Fig. \ref{04} $m_{ee}$ as a
function of the lightest neutrino mass for both mass hierarchies where the
Majorana phases $\alpha$ and $\beta$\ are allowed to vary in the range
$\left[  0-2\pi\right]  $; we find that the $3\sigma$ allowed regions for the
effective Majorana mass are $m_{ee}(\mathrm{eV})\in\lbrack0.00017,0.06084]$,
which corresponds to $m_{1}(\mathrm{eV})\in\lbrack0.00012,0.08267]$ for the
normal hierarchy, and $m_{ee}(\mathrm{eV})\in\lbrack0.02286,0.05878]$, which
corresponds to $m_{3}(\mathrm{eV})\in\lbrack0.00144,0.05879]$ for the inverted
hierarchy. For both hierarchies, the obtained regions of $m_{ee}$ are within
the current experimental data and may be reached in future neutrinoless
double-beta decay experiments \cite{R26}. In particular, the obtained ranges
can be tested in future experiments like KamLAND-Zen, which plans to reach a
sensitivity below $50$ $\mathrm{meV}$ on $\left\vert m_{ee}\right\vert $, and
thus, it will start to constrain the inverted mass hierarchy region \cite{KZ}.

\subsection{Tritium beta decay}

The tritium beta decay is the most sensitive direct way to measure the
absolute neutrino mass scale ignoring the nature of neutrinos \cite{R27}. The
limit (at 95\% C.L.) from the Troitsk and Mainz experiments of the
effective\textrm{\ }electron neutrino mass are, respectively, given by
$m_{\beta}<2.12$ $\mathrm{eV}$ and $m_{\beta}<2.3$ $\mathrm{eV}$
\cite{R28,R29}, while\textrm{\ }the current generation of neutrino mass
measurement comes from the KATRIN experiment with a sensitivity of $m_{\beta
}<0.2$ $\mathrm{eV}$ (at 90 \% C.L.)\textrm{\ }\cite{R30}\textrm{. }The
quantity\textrm{\ }$m_{\beta}$ (or $m_{\nu_{e}}$) is defined in terms of the
mass eigenvalues $m_{i}$ and mixing matrix elements $U_{ei}$:\textrm{\ }%
$m_{\beta}^{2}=\sum\nolimits_{i=1}^{3}U_{ei}^{2}\cdot m_{i}^{2}$.\textrm{ }In
terms of our model parameters, it is expressed as
\begin{equation}
m_{\beta}=\left(  \frac{2m_{1}^{2}}{3}+\frac{m_{2}^{2}}{3}+\frac{m_{3}^{2}}%
{8}\frac{\mathrm{\varepsilon}^{2}}{\left(  a-c\right)  ^{2}}\right)
^{\frac{1}{2}}.
\end{equation}
\begin{figure}[ptbh]
\begin{center}
\hspace{2.5em}\includegraphics[scale=0.56]{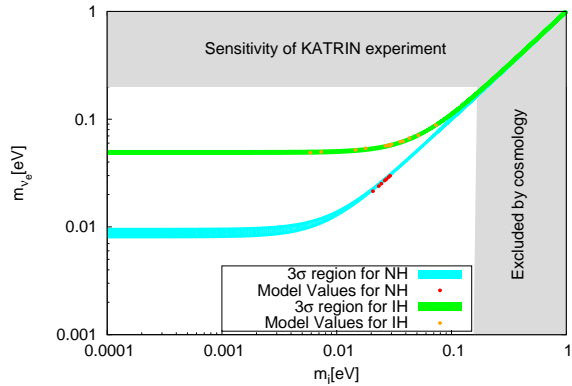}
\end{center}
\caption{$m_{\beta}$\ as a function of the lightest neutrino mass $m_{i}$ for
both mass hierarchies.}%
\label{05}%
\end{figure}Similar to the discussion of the effective Majorana neutrino mass,
in the NH (IH) case, we use the same definitions for $m_{2}$ and $m_{3}$
($m_{1}$ and $m_{2}$). Then, we plot in Fig. \ref{05} the effective\textrm{\ }%
electron neutrino mass $m_{\beta}$\ as a function of the lightest neutrino
mass $m_{i}$. The cyan region (green region) is obtained by varying all the
input parameters\ in their\textrm{\ }$3\sigma$\ ranges for NH (IH) while our
model values are presented by the orange points (the red points). Hence, we
find that the effective\textrm{\ }electron neutrino mass lies in the
range\textrm{\ }$0.0214\lesssim m_{\beta}(\mathrm{eV)}\lesssim0.0298$ for NH
and $0.0488\lesssim m_{\beta}(\mathrm{eV)}\lesssim0.0882$ for IH, while their
corresponding lightest neutrino masses are constrained in the range
$0.0206\lesssim m_{1}(\mathrm{eV})\lesssim0.0291$ for NH and $0.0058\lesssim
m_{3}(\mathrm{eV})\lesssim0.0729$ for IH. The extracted ranges of $m_{\beta}$
are compatible with the above mentioned experiments for both mass hierarchies.
However, the expected future sensitivity from Project 8 \cite{R31} is as low
as $0.04$ $\mathrm{eV}$, which means that only the range corresponding to NH
is allowed.

\subsection{Sum of neutrino masses}

Although the absolute mass scale of the neutrinos remains unknown, the sum of
the three light neutrino masses $\sum\nolimits_{i=1}^{3}\left\vert
m_{i}\right\vert $ is constrained by a cosmological upper bound given by the
Planck Collaboration's limit $\sum\nolimits_{i=1}^{3}\left\vert m_{i}%
\right\vert <0.17$ $\mathrm{eV}$\ \cite{R32}.\textrm{ }In our model, the sum
of neutrino masses is expressed in terms of the model parameters as%
\begin{equation}
m_{\Sigma}=\sum\nolimits_{i=1}^{3}\left\vert m_{i}\right\vert =m_{0}\left(
\mathrm{\varepsilon}-2c-a\right)  .
\end{equation}
\begin{figure}[ptbh]
\begin{center}
\hspace{2.5em}\includegraphics[scale=0.56]{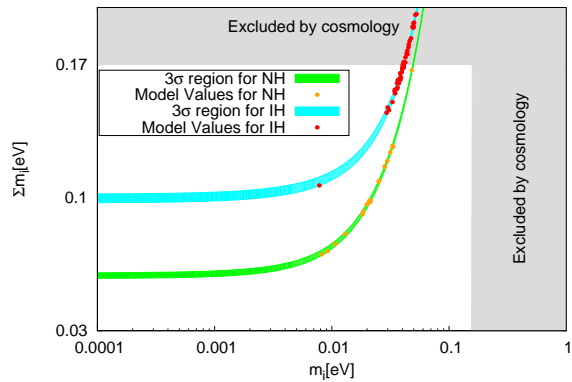}
\end{center}
\caption{Sum of neutrino masses as a function of the lightest neutrino mass
$m_{i}$, present in both cases NH and IH, and the horizontal and vertical
bands in gray correspond to the bounds excluded by cosmology.}%
\label{06}%
\end{figure}Using the $3\sigma$ ranges of mass-squared differences (\ref{dv})
and mixing angles (\ref{mi}), we show in Fig. \ref{06} the dependence of the
sum of the light neutrino masses as a function of the lightest neutrino mass
for both mass hierarchies. The green region (cyan region) is obtained by
varying all the input parameters\ in their\textrm{\ }$3\sigma$\ ranges for
normal hierarchy (inverted hierarchy) while our model values are presented by
the orange points (the red points)\textrm{.} Hence we find that the sum of the
light neutrino masses lies in the range\textrm{\ }$0.0702\lesssim m_{\Sigma
}(\mathrm{eV})\lesssim0.1670$ for NH and $0.1064\lesssim m_{\Sigma
}(\mathrm{eV})\lesssim0.1698$ for IH, while their corresponding lightest
neutrino masses are constrained in the range $0.0081\lesssim m_{1}%
(\mathrm{eV})\lesssim0.0480$ for NH and $0.0078\lesssim m_{3}(\mathrm{eV}%
)\lesssim0.0406$ for IH. Thus, for both mass hierarchies, the sum of neutrino
masses gets more restricted as compared to the Planck limit, and these ranges
may be tested in future cosmological observations.

\subsection{Dirac $CP$ violation}

The Dirac CPV phase $\delta_{CP}$\ is one among the unknown quantities in the
physics of neutrino, and its measurement becomes more important when recent
experiments reported the nonzero value of the reactor angle $\theta_{13}$ as
they are related in the PMNS matrix. Moreover, estimations on the CPV
phase\ $\delta_{CP}$ can be obtained by considering the Jarlskog invariant
quantity $J_{CP}$ which is defined as $J_{CP}=\operatorname{Im}\{U_{\mu
3}U_{e3}^{\ast}U_{e3}U_{\mu3}^{\ast}\}$ and by using the PMNS matrix. It is
expressed as \cite{R33}
\begin{equation}
J_{CP}=\cos\theta_{12}\sin\theta_{12}\cos\theta_{23}\sin\theta_{23}\cos
\theta_{13}^{2}\sin\theta_{13}\sin\delta_{CP}, \label{JC}%
\end{equation}
where the allowed ranges at $3\sigma$ of $\sin\theta_{12},$ $\sin\theta_{23}$,
and $\sin\theta_{13}$ are given in Eq. (\ref{mi}) while the allowed $3\sigma$
ranges of CPV phase $\delta_{CP}$ are giving by \cite{R8}%
\begin{equation}
0\leq\delta_{CP}\leq2\pi\text{ \ for NH \ \ \ },\text{ \ \ \ }0.8\pi\leq
\delta_{CP}\leq2.17\pi\text{ \ for IH.}%
\end{equation}
\begin{figure}[th]
\begin{center}
\hspace{2.5em} \includegraphics[width=.44\textwidth]{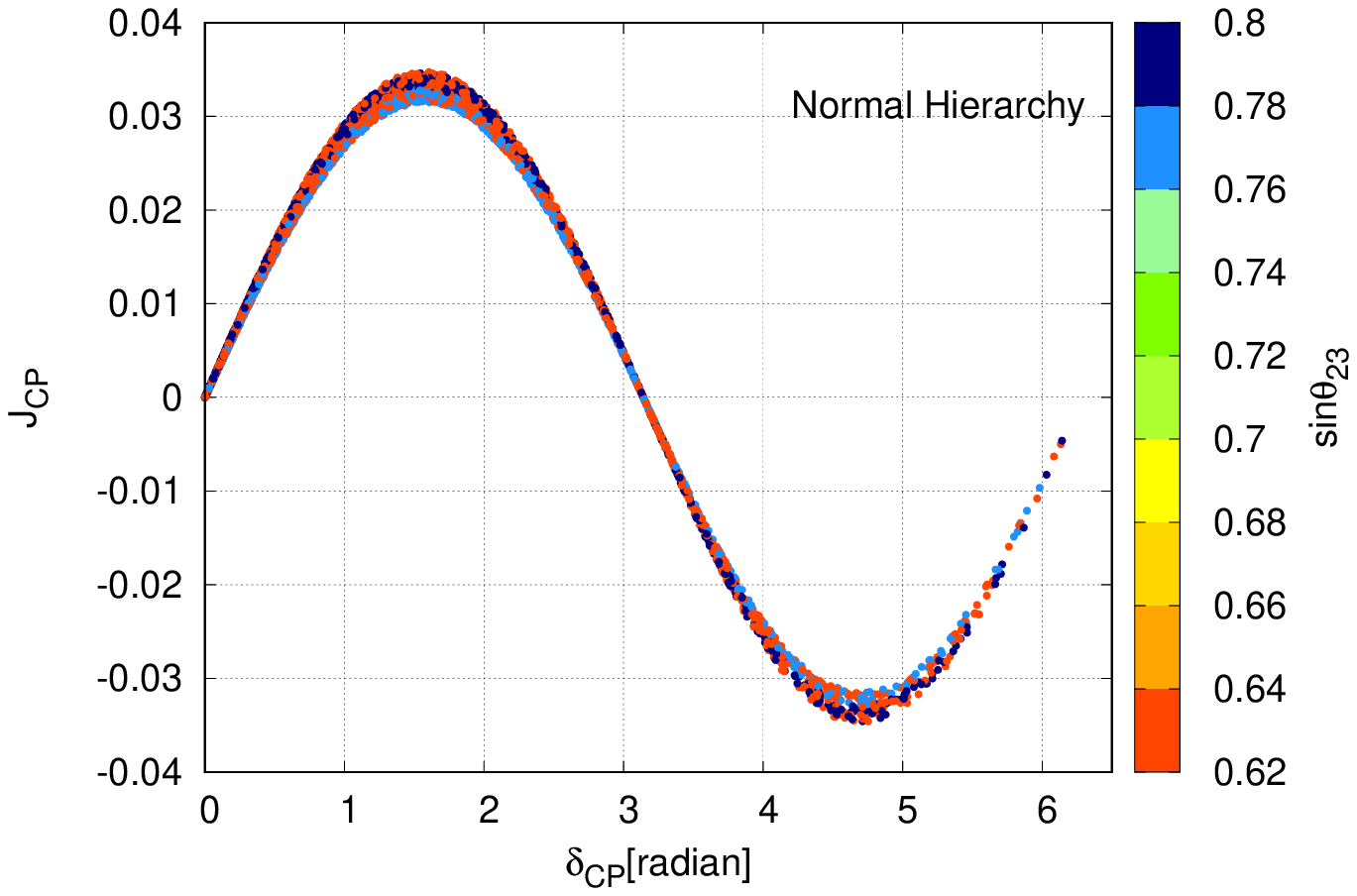}\quad
\includegraphics[width=.44\textwidth]{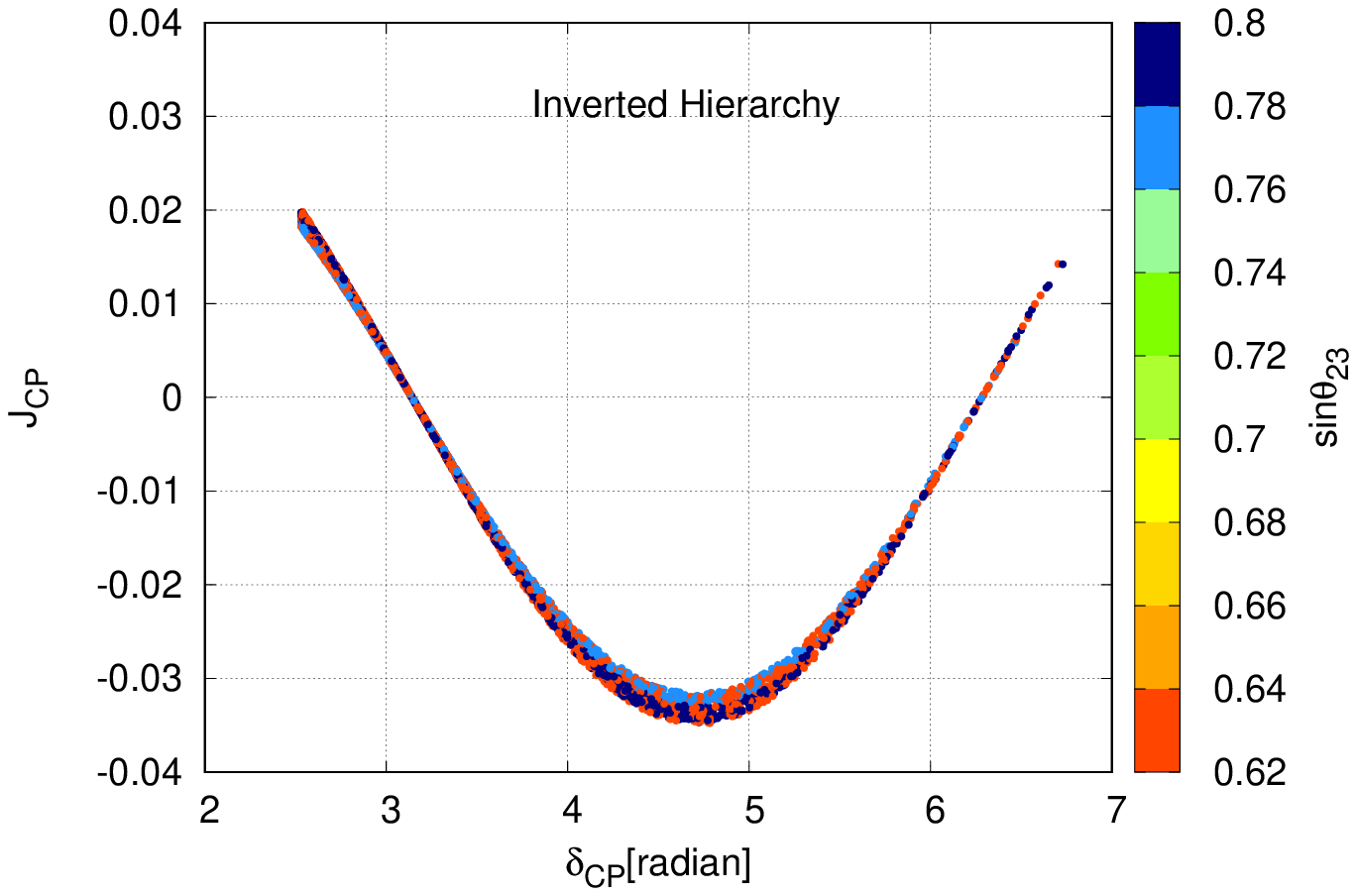}
\end{center}
\par
\vspace{0.5cm}\caption{Left: Scatter plot of $\sin\theta_{23}$ in the
($J_{CP},\delta_{CP}$) plane for normal hierarchy. Right: Same as in left
panel for IH.}%
\label{07}%
\end{figure}We show in Fig. \ref{07}, the behavior of $J_{CP}$\ as a function
of\ $\delta_{CP}$\ with $\sin\theta_{23}$ presented in the palettes for NH
(left panel) and IH (right panel). The ranges of $J_{CP}$ and their
corresponding Dirac CPV $\delta_{CP}$ as well as the ranges of $\sin
\theta_{23}$\ for both HO and LO are as shown in Table \ref{t3}.
\begin{table}[h]
\centering\renewcommand{\arraystretch}{1.2} $%
\begin{tabular}
[c]{|c|c|c|c|c|}\hline
& $J_{CP}$ & $\delta_{CP}$ & $\sin\theta_{23}$ & Color regions\\\hline
NH(HO) & $\left[  -0.034,0.034\right]  $ & \multicolumn{1}{|c|}{$\left[
0,1.91\pi\right]  $} & $\left[  0.776,0.784\right]  $ & Blue and dark
blue\\\hline
NH(LO) & $\left[  -0.034,0.034\right]  $ & \multicolumn{1}{|c|}{$\left[
0,1.93\pi\right]  $} & $\left[  0.629,0.637\right]  $ & Dark orange\\\hline
IH(HO) & $\left[  -0.034,0.019\right]  $ & $\left[  0.8\pi-2.08\pi\right]  $ &
$\left[  0.779,0.784\right]  $ & Blue and dark blue\\\hline
IH(LO) & $\left[  -0.034,0.019\right]  $ & $\left[  0.8\pi-2.14\pi\right]  $ &
$\left[  0.629,0.637\right]  $ & Dark orange\\\hline
\end{tabular}
\ $\caption{Allowed ranges of $J_{CP}$ for both mass hierarchies and both
octants and their corresponding $\delta_{CP}$ and $\sin\theta_{23}$ ranges
extracted from Fig. \ref{07}.}%
\label{t3}%
\end{table}Therefore, the left panel shows that for the values around
$\delta_{CP}=$\ $0.5\pi$ and $\delta_{CP}=$\ $1.5\pi,$ the $CP$ is maximally
violated when the magnitude of $J_{CP}$ is maximal ($J_{CP}\approxeq-0.034$
and $J_{CP}\approxeq0.034$) while in the right panel it is maximally violated
($J_{CP}\approxeq-0.034$ and $J_{CP}\approxeq0.019$) around the values
$\delta_{CP}=0.8\pi$ and $\delta_{CP}=$\ $1.5\pi$.

\section{Conclusion}

In this work, we have constructed a renormalizable hybrid seesaw neutrino
model in the framework of SUSY $SU(5)$ GUT extended by a discrete
$\mathbb{A}_{4}$ family symmetry. The dominant TBM pattern is obtained from
Type I seesaw mechanism while Type II seesaw is responsible for a small
deviation from TBM. Both seesaws are controlled by the action of the
$\mathbb{A}_{4}$ flavor symmetry through its algebraic properties. We found
that the predictions of our proposal concerning the mixing angles and masses
are consistent with the recent measurements.\textrm{ }In particular, we showed
that the deviation by Type II seesaw leads to a nonmaximal atmospheric angle
$\theta_{23}$ as reported recently by the NOvA experiment\ and a nonvanishing
reactor angle $\theta_{13}$. Thus, we made a full analysis depending on the
octant of $\theta_{23}$.\newline We also studied\ the phenomenological
consequences of our proposal where we showed through scatter plots the allowed
ranges for the physical observables and model parameters which we have
restricted by using the $3\sigma$ ranges\ of the neutrino oscillation
parameters for both mass hierarchies. We found also that the sum of neutrino
masses and CPV phase are within the allowed experimental regions. Furthermore,
we found that the ranges of the physical observables involving the\ effective
Majorana\textrm{\ }neutrino mass $m_{\beta\beta}$ and the electron neutrino
mass $m_{\beta}$ are preferred in the case of normal mass hierarchy. For the
latter, the obtained range of $m_{\beta}$ in the inverted mass hierarchy case
is forbidden\textrm{\ }by future sensitivity from Project 8.

\section*{APPENDIX A: CHARGED FERMION SECTORS AND PROTON DECAY}

In this appendix, we provide a brief study of the charged lepton sector to
show the possibility to use an $\mathbb{A}_{4}$ assignment for the remaining
$SU(5)$ superfield content that does not affect the neutrino mixing. However,
it is well known in GUTs that because the quarks and leptons are unified in
the same group representations, the charged lepton and the down quark masses
are derived from the same superpotential. Thus we also provide in this
appendix a concise discussion of the quark sector fixing up the unwanted mass
relations between down quarks and charged leptons
\begin{equation}
m_{e}=m_{d}\quad,\quad m_{\mu}=m_{s}\quad,\quad m_{\tau}=m_{b}. \tag{A.1}%
\label{de}%
\end{equation}

We begin by assigning the quantum numbers to the rest of the chiral
superfields of our $SU(5)\times\mathbb{A}_{4}$ proposal. Thus, in addition to
the superfields relevant for the neutrino sector---see Table \ref{t1}---the
matter $10_{m}^{i}=(U_{i}^{c},E_{i}^{c},Q_{i})$ of the three generations
$\mathrm{i=1,2,3}$\ live in the $\mathbb{A}_{4}$\ representations $1_{\left(
1,\omega\right)  }$, $1_{\left(  1,\omega^{2}\right)  }$, and $1_{\left(
1,1\right)  }$, respectively. As discussed in the neutrino sector above,\ one
flavon superfield is\ necessary to accommodate the observed neutrino
oscillation parameters. Similarly, to generate appropriate masses for the
three generations of up quarks and down quarks (as well as charged leptons),
two extra flavons are needed in the superpotential of up quarks $W_{u}$; these
are denoted by $\chi$ and\ $\varphi$. On the other hand, three extra flavons
are required in the superpotential of down quarks and charged leptons
$W_{e,d}$; these are denoted as $\rho,$ $\eta$, and $\sigma$. The
$\mathbb{A}_{4}$ irreducible representations of these new flavons are as given
in Table \ref{t4}.\begin{table}[h]
\centering \renewcommand{\arraystretch}{1.2}
\begin{tabular}
[c]{|l|l|l|l|l|l|l|}\hline
Flavons & $\Phi$ & $\chi$ & $\varphi$ & $\rho$ & $\sigma$ & $\eta$\\\hline
$\mathrm{SU(5)}$ & $1$ & $1$ & $1$ & $1$ & $1$ & $1$\\\hline
$\mathbb{A}_{4}$ & $3_{-1,0}$ & $1_{\left(  1,\omega\right)  }$ & $1_{\left(
1,\omega^{2}\right)  }$ & $3_{-1,0}$ & $3_{-1,0}$ & $3_{-1,0}$\\\hline
$\mathrm{U(1)}$ & $0$ & $-4$ & $2$ & $7$ & $8$ & $5$\\\hline
\end{tabular}
\caption{Flavon superfields needed in the quark and charged lepton sectors and
their quantum numbers under $SU(5)\times\mathbb{A}_{4}\times U(1)$.}%
\label{t4}%
\end{table}Furthermore, in order to achieve the correct mass hierarchy and to
get rid of the unwanted couplings, we add an additional global $U(1)$ symmetry
where its charge assignments for all the superfields in our proposal are as
given in the last rows of Tables \ref{t4} and \ref{t5}. In fact, these $U(1)$
quantum numbers are identified by taking into account the preexisting
$SU(5)\times\mathbb{A}_{4}$ invariant Yukawa couplings in the neutrino sector.
Indeed, the flavon $\Phi$\ must carry a zero $U(1)$ charge in order to
preserve both couplings given in the Majorana superpotential (\ref{wm}).
However, since the nonrenormalizable terms up to order $\mathcal{O}%
(1/\Lambda^{2})$ are needed in the charged fermion sectors as we will see
below, this zero $U(1)$ charge for the flavon $\Phi$ enables its coupling with
the operators $F_{i}F_{i}H_{\overline{15}}$ and $N_{i}F_{i}H_{5}$ via the
following higher dimensional operators:%
\begin{equation}
F_{i}F_{i}H_{\overline{15}}\left(  \frac{\Phi}{\Lambda}\right)  \quad,\quad
N_{i}F_{i}H_{5}\left(  \frac{\Phi}{\Lambda}\right)  . \tag{A.2}\label{ff}%
\end{equation}
These couplings which destroy the form of neutrino mass matrix (\ref{gm}) that
led to the desired oscillation parameters must be suppressed. This is possible
if we assume that $\upsilon_{\Phi}\ll\Lambda,$ which is acceptable according
to Eqs. (\ref{maj}) and (\ref{pp}). On the other hand, even if the VEV of the
flavon $\Phi$\ is around the cutoff scale---say $\Phi\simeq\Lambda$---this
would just give terms that are relative to the leading ones: $F_{i}%
F_{i}H_{\overline{15}}$\ and $N_{i}F_{i}H_{5}$. \textrm{\ }\begin{table}[h]
\centering \renewcommand{\arraystretch}{1.2}
\begin{tabular}
[c]{|l|l|l|l|l|l|l|l|l|l|}\hline
Fields & $F_{i}$ & $N_{i}$ & $H_{5}$ & $H_{\bar{5}}$ & $H_{15}$ & $T_{1}$ &
$T_{2}$ & $T_{3}$ & $H_{\overline{45}}$\\\hline
$\mathrm{SU(5)}$ & $\bar{5}_{m}^{i}$ & $1_{\nu}^{i}$ & $5_{H_{u}}$ &
$5_{H_{d}}$ & $15_{\Delta_{d}}$ & $10_{m}^{1}$ & $10_{m}^{2}$ & $10_{m}^{3}$ &
$\overline{45}_{H}$\\\hline
$\mathbb{A}_{4}$ & $3_{-1,0}$ & $3_{-1,0}$ & $1_{\left(  1,1\right)  }$ &
$1_{\left(  1,\omega\right)  }$ & $1_{\left(  1,\omega\right)  }$ &
$1_{\left(  1,\omega\right)  }$ & $1_{\left(  1,\omega^{2}\right)  }$ &
$1_{\left(  1,1\right)  }$ & $3_{-1,0}$\\\hline
$\mathrm{U(1)}$ & $-2$ & $0$ & $2$ & $-4$ & $4$ & $1$ & $-2$ & $-1$ &
$4$\\\hline
\end{tabular}
\caption{Matter and Higgs content of the model and their quantum numbers under
$SU(5)\times\mathbb{A}_{4}\times U(1)$.}%
\label{t5}%
\end{table}Moreover, it is well known that the $SU(5)$ GUT predicts the mass
relations in Eq. (\ref{de}), which are not acceptable for the first and second
generations due to their disagreement with the experimental data
\cite{B4}.\textrm{\ }Nevertheless, the well known Georgi-Jarlskog (GJ)
mechanism \cite{GJ}\textrm{\ }overcomes this issue by introducing an
additional Higgs in the $45$-dimensional $SU(5)$ representation leading to the
mass relations
\begin{equation}
3m_{e}=m_{d\quad},\quad m_{\mu}=3m_{s}\quad,\quad m_{\tau}=m_{b}.
\tag{A.3}\label{GJ}%
\end{equation}
In our proposal, this $45$ Higgs denoted as $H_{\overline{45}}$ is placed in
an $\mathbb{A}_{4}$\ triplet\footnote{Notice that the choice of putting the
45-dimensional Higgs $H_{\overline{45}}$ in an $\mathbb{A}_{4}$ triplet is to
ensure the invariance of its coupling with $T_{2}F_{i}$ without having to add
other flavon triplets.}---$\tilde{H}_{\overline{45}}=(H_{\overline{45}%
},0,0)^{T}$---while its charge under the additional $U(1)$ symmetry is
$q_{U(1)}=-5$. Recall that this Higgs $H_{\overline{45}}$\ is antisymmetric in
$SU(5)$ indices and satisfies the following relations \cite{GJ}:
\begin{equation}%
\begin{array}
[c]{ccccc}%
(H_{\overline{45}})_{c}^{ab} & = & -(H_{\overline{45}})_{c}^{ba} & , &
(H_{\overline{45}})_{a}^{ab}=0,\\
\left\langle (H_{\overline{45}})_{i}^{i5}\right\rangle  & = & \upsilon_{45} &
, & i=1,2,3,\\
\left\langle (H_{\overline{45}})_{4}^{45}\right\rangle  & = & -3\upsilon
_{45}. &  &
\end{array}
\tag{A.4}\label{H45}%
\end{equation}
With the $\mathbb{A}_{4}\times U(1)$ charge assignments shown in Table
\ref{t5}, the usual renormalizable Yukawa couplings $Y_{1}T_{1}F_{i}H_{\bar
{5}}$, $Y_{2}T_{2}F_{i}H_{\bar{5}}$,\ and $Y_{3}T_{3}F_{i}H_{\bar{5}}$\ are
not invariant under $\mathbb{A}_{4}$\ flavor symmetry and they are carrying
the $U(1)$\ charges $-5$, $-8$,\ and $-7$, respectively. Thus, to restore the
invariance under the $\mathbb{A}_{4}\times U(1)$ symmetry,\ each one of these
couplings requires a different $\mathbb{A}_{4}$\ triplet flavon
superfield,\ namely $\eta$, $\sigma$,\ and $\rho$\ with $U(1)$\ charges $5,$
$8$,\ and $7$, respectively. Therefore, the $\mathbb{A}_{4}\times U(1)$
invariant superpotential of the down quarks and charged leptons involving the
three flavons $\eta$, $\sigma$,\ and $\rho$\ as well the\ Higgs\textrm{\ }%
$H_{\overline{45}}$\ is given by%
\begin{equation}
W_{d,e}=\frac{Y_{1}}{\Lambda}T_{1}\left(  F_{i}\eta\right)  H_{\bar{5}}%
+\frac{Y_{2}}{\Lambda}T_{2}\left(  F_{i}\sigma\right)  H_{\bar{5}}+\frac
{Y_{3}}{\Lambda}T_{3}\left(  F_{i}\rho\right)  H_{\bar{5}}+Y_{45}T_{2}%
F_{i}\tilde{H}_{\overline{45}}, \tag{A.5}\label{wd}%
\end{equation}
where $Y_{1}$, $Y_{2}$, $Y_{3}$,\ and $Y_{45}$ are the Yukawa mass matrices
and $\Lambda$ represents the cutoff scale of the model. Notice that the
coupling $T_{2}F_{i}\tilde{H}_{\overline{45}}\left(  \frac{\Phi}{\Lambda
}\right)  $\ is also allowed by the symmetries of the model, but again its
suppression is guaranteed by the condition\textrm{\ }$\upsilon_{\Phi}%
\ll\Lambda$. Using $\mathbb{A}_{4}$ tensor products, the superpotential
$W_{d,e}$\ develops into
\begin{equation}
W_{d,e}=\frac{Y_{1}}{\Lambda}T_{1}F_{2}\eta H_{\bar{5}}+\frac{Y_{2}}{\Lambda
}T_{2}F_{1}\sigma H_{\bar{5}}+\frac{Y_{3}}{\Lambda}T_{3}\left(  F_{3}%
\rho\right)  H_{\bar{5}}+Y_{45}T_{2}F_{2}H_{\overline{45}}. \tag{A.6}%
\label{wde}%
\end{equation}
The masses arise from the breaking of $\mathbb{A}_{4}\times U(1)$ family
symmetry as well as the breaking of the electroweak symmetry. Therefore, by
taking the flavon triplet VEVs along the directions~%
\begin{equation}
\left\langle \sigma\right\rangle =\upsilon_{\sigma}(1,0,0)^{T}\quad
,\quad\left\langle \rho\right\rangle =\upsilon_{\rho}(1,0,0)^{T}\quad
,\quad\left\langle \eta\right\rangle =\upsilon_{\eta}(1,0,0)^{T}, \tag{A.7}%
\end{equation}
the Higgs doublet $H_{d}$ responsible for the electroweak symmetry breaking as
usual $\left\langle H_{d}\right\rangle =\upsilon_{d}$, and the Higgs $45$ as
in Eq. (\ref{H45}), we obtain the mass matrices for down-type quarks $M_{d}$
and charged leptons $M_{e}$%

\begin{equation}
M_{d}=\left(
\begin{array}
[c]{ccc}%
0 & Y_{1}r & 0\\
Y_{2}h & Y_{45}\upsilon_{45} & 0\\
0 & 0 & Y_{3}t
\end{array}
\right)  \text{ \ },\text{ \ }M_{e}=\left(
\begin{array}
[c]{ccc}%
0 & Y_{2}h & 0\\
Y_{1}r & -3Y_{45}\upsilon_{45} & 0\\
0 & 0 & Y_{3}t
\end{array}
\right)  , \tag{A.8}\label{mde}%
\end{equation}
$\allowbreak$where $r=\upsilon_{d}\upsilon_{\eta}/\Lambda$, $h=\upsilon
_{d}\upsilon_{\sigma}/\Lambda$, and $t=\upsilon_{d}\upsilon_{\rho}/\Lambda$.
By assuming $Y_{45}\upsilon_{45}\gg Y_{1}r\approx Y_{2}h$, we diagonalize the
mass matrices $M_{d}$ and $M_{e}$ where we find that the masses of down-type
quarks and charged leptons are respectively given by%
\begin{align}
m_{d}  &  =\left\vert \frac{Y_{1}^{2}}{Y_{45}}\frac{r^{2}}{\upsilon_{45}%
}\right\vert \quad,\quad m_{s}=\left\vert Y_{45}\upsilon_{45}+\frac{Y_{1}^{2}%
}{Y_{45}}\frac{r^{2}}{\upsilon_{45}}\right\vert \quad,\quad m_{\tau
}=\left\vert Y_{3}t\right\vert ,\nonumber\\
m_{e}  &  =\left\vert \frac{Y_{1}^{2}}{3Y_{45}}\frac{r^{2}}{\upsilon_{45}%
}\right\vert \quad,\quad m_{\mu}=\left\vert 3Y_{45}\upsilon_{45}+\frac
{Y_{1}^{2}}{3Y_{45}}\frac{r^{2}}{\upsilon_{45}}\right\vert \quad,\quad
m_{b}=\left\vert Y_{3}t\right\vert , \tag{A.9}\label{mqd}%
\end{align}
where these masses imply the Georgi-Jarlskog relations given in Eq.
(\ref{GJ}). Notice that these mass relations are admissible at the GUT scale
at leading order and can be improved assuming the SUSY threshold corrections
and appropriate values of $\tan\beta=\frac{\upsilon_{u}}{\upsilon_{d}}$; for
more details on the SUSY threshold corrections procedure see Refs. \cite{GJ1,
TH}. On the other hand, an alternative way to go beyond the $b-\tau$
unification in GJ predictions at high scale is through higher dimensional
effective operators \cite{GJ1, GJ2}. These operators involve additional
Higgses in $24_{H}$ or $75_{H}$ and a nontrivial $SU(5)$ messenger fields $X$
and $\overline{X}$ allowing for relations such as $m_{\tau}=\frac{3}{2}m_{b}$.
All possible relations between down-quark and the charged lepton masses are
listed in Table 1 of Ref. \cite{GJ1} and Table 2 of Ref. \cite{GJ2}. One of
these GUT scale relations using fermion and scalar messenger fields is studied
in the framework of $SU(5)\times\mathbb{A}_{4}$ in Ref. \cite{GJ3}.

Regarding the up-type quark sector, besides the top quark mass which is
preferred to arise from a renormalizable coupling, the remaining up and charm
quark masses are derived from higher dimensional Yukawa couplings involving
flavon superfields. Indeed, in our model, two different flavons $\chi$ and
$\varphi$ couple to the first and second generations, respectively. Thus, the
superpotential of the up-type quarks respecting gauge and flavor symmetries
takes the form
\begin{equation}
W_{u}=\frac{Y^{u}}{\Lambda}T_{1}T_{1}H_{5}\chi+\frac{Y^{c}}{\Lambda}T_{2}%
T_{2}H_{5}\varphi+Y^{t}T_{3}T_{3}H_{5}, \tag{A.10}\label{wu}%
\end{equation}
where $y^{u}$, $y^{c}$,\ and $y^{t}$\ are the Yukawa coupling constants for
up-, charm-, and top-type quarks. As usual, the up-type quark masses arise
from the breaking of the flavor and electroweak symmetries. Thus, when the
flavons $\varphi$ and $\chi$ and the Higgs $H_{u}$ develop their VEVs as%
\begin{equation}
\left\langle \varphi\right\rangle =\upsilon_{\varphi}\quad,\quad\left\langle
\chi\right\rangle =\upsilon_{\chi}\quad,\quad\left\langle H_{u}\right\rangle
=\upsilon_{u}, \tag{A.11}%
\end{equation}
we obtain a diagonal mass matrix of the up-type quarks given by%
\begin{equation}
M_{up}=\upsilon_{u}\left(
\begin{array}
[c]{ccc}%
Y^{u}\upsilon_{\chi}/\Lambda & 0 & 0\\
0 & Y^{c}\upsilon_{\varphi}/\Lambda & 0\\
0 & 0 & Y^{t}%
\end{array}
\right)  \tag{A.12}\label{up}%
\end{equation}
with the mass eigenvalues as%
\begin{equation}
m_{u}=Y^{u}\frac{\upsilon_{\chi}}{\Lambda}\upsilon_{u}\quad,\quad m_{c}%
=Y^{u}\frac{\upsilon_{\varphi}}{\Lambda}\upsilon_{u}\quad,\quad m_{t}%
=Y^{t}\upsilon_{u}. \tag{A.13}%
\end{equation}
The large mass of the top quark\textrm{\ }is\ obtained at tree level, while
the mass hierarchy among the first two generations of up-type quarks can be
obtained by assuming a hierarchy between the VEVs of the flavons $\chi$\ and
$\varphi$.\newline As for the mixing in the quark sector, it is defined as
$\left\vert U_{Q}\right\vert =\left\vert U_{up}^{\dagger}U_{d}\right\vert $
where $U_{d}$ is the matrix that diagonalizes the mass matrix of down quarks
$M_{d}$\ while $U_{up}$ is the one that diagonalizes the mass matrix of up
quarks $M_{up}$. Since this latter is diagonal (\ref{up}), $U_{up}$ is just
the identity matrix, and thus, the total mixing matrix is the one that
diagonalizes the mass matrix of the down quarks $M_{d}$\ (\ref{mde}); we find%
\begin{equation}
U_{Q}=U_{d}=\left(
\begin{array}
[c]{ccc}%
\frac{-Y_{45}\upsilon_{45}-F}{ZY_{1}r} & \frac{-Y_{45}\upsilon_{45}+F}%
{EY_{1}r} & 0\\
\frac{2}{Z} & \frac{2}{E} & 0\\
0 & 0 & 1
\end{array}
\right)  \tag{A.14}\label{uq}%
\end{equation}
with%
\begin{equation}
F=\sqrt{Y_{45}^{2}\upsilon_{45}^{2}+4Y_{1}^{2}r^{2}},~Z=\sqrt{4+\left(
\frac{Y_{45}\upsilon_{45}+F}{Y_{1}r}\right)  ^{2}},~E=\sqrt{4+\left(
\frac{Y_{45}\upsilon_{45}-F}{Y_{1}r}\right)  ^{2}}. \tag{A.15}%
\end{equation}
Notice that the zero entries in the mixing matrix (\ref{uq}) can be seen to be
a first approximation to the mixing matrix $V_{CKM}$ of the quark sector
\cite{B4}. The nonzero values of this entries can be obtained by considering
higher dimensional operators involving flavon superfields in the quark sector.
As for\textrm{\ }the mixing in the charged lepton sector, the diagonalization
of the mass matrix $M_{e}$\ in Eq. (\ref{mde}) is given by%

\begin{equation}
U_{e}\simeq\left(
\begin{array}
[c]{ccc}%
\frac{3Y_{45}\upsilon_{45}-L}{GY_{1}r} & \frac{3Y_{45}\upsilon_{45}+L}%
{KY_{1}r} & 0\\
\frac{2}{G} & \frac{2}{K} & 0\\
0 & 0 & 1
\end{array}
\right)  \tag{A.16}%
\end{equation}
with%
\begin{equation}
L=\sqrt{9Y_{45}^{2}\upsilon_{45}^{2}+4Y_{1}^{2}r^{2}},~G=\sqrt{4+\left(
\frac{-3Y_{45}\upsilon_{45}+L}{Y_{1}r}\right)  ^{2}},~K=\sqrt{4+\left(
\frac{3Y_{45}\upsilon_{45}+L}{Y_{1}r}\right)  ^{2}}. \tag{A.17}%
\end{equation}
From this matrix, it is clear that the charged lepton mixing angles
$\theta_{13}^{l}$ and $\theta_{23}^{l}$\ are both equal to zero; thus in our
model the mixing from the charged lepton sector does not affect the mixing
angles of the neutrino sector given in Eq. (\ref{ss}). Notice by the way that
the total mixing in the lepton sector $U_{\mathrm{PMNS}}=U_{e}^{\dagger}%
\tilde{U}$ is proportional to $\tilde{U}$ with a small shift of the mixing
angle $\theta_{12}^{l}$.

We end this appendix by giving comments concerning the well-known four- and
five-dimensional operators that contribute to fast proton decay in
supersymmetric $SU(5)$ GUT models. In this respect, the dangerous proton decay
terms arise from the dimension four $\lambda^{ijk}10_{m}^{i}\bar{5}_{m}%
^{j}\bar{5}_{m}^{k}$\ and dimension five $\lambda^{ij}\lambda^{kl}10_{m}%
^{i}10_{m}^{j}10_{m}^{k}\bar{5}_{m}^{l}$ operators.\emph{\ }These\emph{\ }%
operators are dangerous in the sense that they lead to proton decay rates far
larger than the experimental limits. As regards to the former operators, they
contribute to the proton decay through the term violating baryon number
$(U_{1}^{c}D_{1}^{c}D_{k}^{c})$\ combined with the term $(Q_{i}L_{j}D_{k}%
^{c})$\ that violates the lepton number with family indices as $i,$ $j=1,$ $2$
and $k=2,$ $3$. In fact, these operators which are renormalizable can be
avoided by imposing the usual $R$ parity as in the case of the MSSM \cite{RP}.
However, in our $SU(5)\times\mathbb{A}_{4}\times U(1)$ proposal, these
four-dimensional operators that are given by
\begin{equation}
10_{m}^{i}\bar{5}_{m}^{j}\bar{5}_{m}^{j}\rightarrow T_{1}F_{j}F_{j}+T_{2}%
F_{j}F_{j}+T_{3}F_{j}F_{j}\tag{A.18}%
\end{equation}
are prevented by the additional $U(1)$ symmetry. On the other hand, in flavor
symmetries based models there are additional nonrenormalizable couplings which
involve flavon fields and can generate proton decay operators. In our model,
these nonrenormalizable operators up to order $\mathcal{O}(1/\Lambda^{2})$
look like $\left(  1/\Lambda\right)  10_{m}^{i}\bar{5}_{m}^{j}\bar{5}_{m}%
^{j}\Omega$ with $\Omega=\varphi,\chi,\rho,\sigma,\eta$ as the various flavon
superfields used throughout the different sectors studied in this work. It is
easy to check from Tables \ref{t4} and \ref{t5} that these couplings are also
not allowed as they are not invariant under the $U(1)$ symmetry.

Regarding the five-dimensional couplings $\lambda^{ijkl}10_{m}^{i}10_{m}%
^{j}10_{m}^{k}\bar{5}_{m}^{l}$, they are mediated by the heavy color triplet
Higgsino and it is well known that their dressing diagrams\footnote{The
dressing procedure of five-dimensional operators consists of converting two
scalars (sfermions) in the $TTTF$ couplings to two fermions by a loop diagram
through the exchange of winos and Higgsinos---these are the dominant
contributions to the operators $qqql$ and $u^{c}u^{c}d^{c}e^{c}$,
respectively. For more details and examples on such diagrams see, for
instance, Ref. \cite{DD}.} to form six-dimensional operators are the most
disturbing operators that lead to fast proton decay in SUSY $SU(5)$ models
\cite{DC,Nath}.\textrm{\ }These operators that are derived from the
renormalizable up and down Yukawa couplings $\lambda TTH_{5}$ and
$\lambda^{\prime}TFH_{\bar{5}}$ are absent in our model since they behave,
respectively, as nontrivial singlets and triplet under the $\mathbb{A}_{4}$
flavor symmetry. However, the last couplings---which are required to generate
masses for the charged fermions---are allowed through their interactions with
the flavon superfields as given in the Yukawa couplings (\ref{wu}) and
(\ref{wde}). Thus, our model contains higher order operators of the kind
$\frac{1}{M_{T}}TTTF\left(  \frac{\Omega}{\Lambda}\right)  ^{n}$ where $M_{T}$
is the mass of the colored Higgs triplet and $n=1,2$; for $n=1$ we have
$\Omega=\eta$, and for $n=2$ the relevant combinations are $\Omega^{2}%
=\sigma\chi,\rho\chi,\eta\varphi$. Hence, the suppression of these operators
compared to the usual five-dimensional couplings $TTTF$ is now enhanced by the
factors $\left(  \frac{\Omega}{\Lambda}\right)  ^{n}$ coming from the flavon
superfields required by $\mathbb{A}_{4}$ invariance, thus leading to highly
suppressed proton decay. We should note, however, that to provide precise
predictions for the proton decay rate, the renormalization group equations
(RGEs) for the gauge couplings at one loop must be taken into account
\cite{FF}; this clearly goes beyond the scope of this paper.

\end{document}